\documentclass[preprint]{aastex701}

\usepackage[T1]{fontenc}
\usepackage[utf8]{inputenc}

\usepackage{amsmath}
\usepackage{amssymb}
\usepackage{amsfonts}
\usepackage{amsthm}
\usepackage{mathtools}

\usepackage{bm}
\usepackage{xcolor}
\usepackage[normalem]{ulem}
\usepackage{booktabs}
\usepackage{array}
\usepackage{multirow}
\usepackage{float}
\usepackage{fvextra}

\graphicspath{{}{moon_cross_tidal_figures/}{./}}

\newtheorem{proposition}{Proposition}

\newcommand{\E}{\mathcal{E}}
\newcommand{\A}{\mathcal{A}}
\newcommand{\R}{\mathcal{R}}

\newcommand{\dd}{\mathrm{d}}

\newcommand{\eps}{\varepsilon}

\newcommand{\tr}{\operatorname{tr}}
\newcommand{\atanTwo}{\operatorname{atan2}}

\newcommand{\kM}{\kappa_M}
\newcommand{\RE}{R_E}
\newcommand{\MM}{M_M}

\newcommand{\DHal}{D_H}
\newcommand{\NH}{N_H}
\newcommand{\PH}{P_H}

\newcommand{\etaH}{\eta_H}
\newcommand{\chioff}{\chi_{\rm off}}
\newcommand{\gamH}{\gamma_H}

\newcommand{\bchi}{\bm{\chi}}

\newcommand{\As}{A_s}
\newcommand{\Ac}{A_c}
\newcommand{\Asframe}{A_s^{\rm frame}}
\newcommand{\dframe}{\delta}

\makeatletter

\let\frontmatter@produce@dates\relax

\let\frontmatter@produce@footnote\relax

\makeatother

\renewcommand{\bfseries}{\mdseries}

\begin{document}

\title{
    Frame Degeneracy and Sine-Quadrature Residuals in Lunar Tidal Modeling
}

\correspondingauthor{Muhittin Cenk Eser}

\author{Muhittin Cenk Eser}
\affiliation{
Department of Mathematics and Computer Science,
Freie Universit\"at Berlin,
Arnimallee 6, 14195 Berlin, Germany
}
\email{muhittin.eser@fu-berlin.de}

\date{\today}

\begin{abstract}
The dominant Earth--Moon tide is accurately described by Newtonian theory, but interpreting a possible off-diagonal residual requires separating its signature from errors in the standard tidal model. In a local lunar principal frame, the Newtonian plus channel varies as \(\cos(2\beta)\), whereas a symmetric off-diagonal residual enters the \(\sin(2\beta)\) quadrature. We extend this separation to a globally defined three-dimensional symmetric trace-free tensor with two transverse coefficients projected onto each station vertical. This formulation establishes an exact first-order identifiability limit: lunar-frame errors generate responses indistinguishable from the cross-residual signal, making finite constraints intrinsically prior limited. A joint-prior GLS/Fisher forecast incorporates frame, ephemeris, projection, terrestrial-response, loading, environmental, instrumental, and correlated-network uncertainties. The dimensionless worst-principal-axis uncertainties are \(9.513\times10^{-3}\), \(4.490\times10^{-4}\), and \(4.182\times10^{-5}\) for conservative, moderate, and optimistic synthetic networks. Monte Carlo injection--recovery experiments assess robustness using independent calibration and validation halves. Under the declared response-expanded Stress budget, optimistic formal all-draw coverage falls to \(0.317\), with a false-positive rate of \(0.685\); calibrated robust covariance restores held-out coverage to \(0.949\) and the false-positive rate to \(0.051\). These results remain conditional on the synthetic assumptions. The Halilsoy cross-polarized sector provides only a local morphology template; no detection or constraint on \(\gamma_H\) is claimed. The physical contribution is to establish how a generic lunar cross-residual would enter gravimetric observations and what independent constraints are required to distinguish it from standard tidal-model errors, providing a quantitative foundation for future observational tests.
\end{abstract}


\newpage

\section{Introduction}
\label{sec:introduction}

The lunar tide is one of the most accurately modeled gravitational signals in geophysics. 
At leading order, the Earth--Moon system produces a Newtonian quadrupolar tidal tensor. 
In a local principal frame aligned with the instantaneous Earth--Moon direction, the two-dimensional block relevant for the present discussion is
\begin{equation}
\E_N
=
\kM
\begin{pmatrix}
2 & 0
\\
0 & -1
\end{pmatrix},
\qquad
\kM
=
\frac{G\MM}{D^3},
\label{eq:intro_newtonian_tensor}
\end{equation}
where \(\MM\) is the lunar mass and \(D\) is the Earth--Moon distance. 
This tensor stretches along the Earth--Moon direction and squeezes along a transverse direction. 
Its principal axes are orthogonal and define the standard plus-type Newtonian tidal frame \citep{MurrayDermott1999,Melchior1983,Cartwright1999,Agnew2007,PoissonWill2014}.
This Newtonian description is the leading near-field limit of a relativistic tidal interaction. 
In general relativity, tidal forces are encoded in the electric part of the Weyl tensor, and their propagation can be formulated through covariant wave equations for the Weyl sector \citep{goswami2021tidal}. 
The present work does not test the propagation speed of gravity or model the full relativistic two-body problem; it uses this relativistic viewpoint only to motivate a local residual tidal basis in which off-diagonal, cross-type components can be separated from the dominant Newtonian plus channel.

The present paper does not attempt to replace this standard description. 
Instead, it asks a more restricted and observationally meaningful residual question: after the dominant Newtonian lunar tide, solar tide, solid-Earth response, ocean loading, atmospheric and hydrological effects, instrumental drift, and reference-frame corrections have been modeled, what tensorial and harmonic form would a possible off-diagonal lunar tidal residual have? 
This formulation is essential because a diagonal Newtonian tensor can be projected along any direction, including \(45^\circ\). 

Such a projection is not an independent cross-tidal residual; it is only the value of the same diagonal field along a chosen direction.

The observational setting is therefore precision Earth-tide analysis rather than an isolated two-body acceleration estimate. Classical tide-generating potentials organize lunar and solar forcing into harmonic constituents, while practical gravimetry must also represent elastic and self-gravitational Earth response, ocean and environmental loading, polar motion, site response, calibration, drift, and Earth-orientation transformations \citep{Doodson1921,CartwrightEdden1973,Tamura1987,HartmannWenzel1995,PetitLuzum2010,Agnew2007,BoyHinderer2006,VanCampVauterin2005,Neumeyer2010}. The local sine/cosine algebra below identifies the residual quadrature; the quantitative forecast then embeds it in a three-dimensional station response and a nuisance model that allows standard tide-model errors to enter the same observable subspace.

For a unit direction making an angle \(\beta\) with the modeled Earth--Moon principal axis, the projected Newtonian acceleration at the Earth's surface is
\begin{equation}
a_N(\beta)
=
a_0
\left[
\frac12+\frac32\cos(2\beta)
\right],
\qquad
a_0
=
\frac{G\MM\RE}{D^3}.
\label{eq:intro_newtonian_projection}
\end{equation}
The ordinary Newtonian principal-frame projection therefore occupies the plus-type harmonic channel \(\cos(2\beta)\). 
A symmetric off-diagonal residual would instead enter through the sine-quadrature channel,
\begin{equation}
\boxed{
\begin{aligned}
\text{diagonal plus channel}
&\sim
\cos(2\beta),
\\
\text{off-diagonal residual channel}
&\sim
\sin(2\beta).
\end{aligned}
}
\label{eq:intro_harmonic_separation}
\end{equation}
This algebraic separation is the starting point of the proposed residual framework.

However, the same algebra also shows why the problem is delicate. 
A sine-quadrature coefficient is not uniquely diagnostic of a physical off-diagonal tide. 
A small error in the assumed principal-frame direction, projection geometry, ephemeris phase, local vertical, station orientation, or terrestrial response model converts part of the ordinary diagonal Newtonian tide into an apparent \(\sin(2\beta)\) term. 
If the true principal-frame angle is \(\beta-\dframe\), while the analysis uses the nominal angle \(\beta\), then
\begin{equation}
a_N(\beta-\dframe)
=
a_0
\left[
\frac{1}{2}
+
\frac{3}{2}\cos(2\beta)\cos(2\dframe)
+
\frac{3}{2}\sin(2\beta)\sin(2\dframe)
\right].
\label{eq:intro_frame_shift}
\end{equation}
Thus frame misregistration alone produces
\begin{equation}
\Asframe
=
\frac{3}{2}a_0\sin(2\dframe),
\qquad
\chi_{\rm app}
=
\frac{\Asframe}{a_0}
=
\frac{3}{2}\sin(2\dframe)
\simeq
3\dframe .
\label{eq:intro_frame_degeneracy}
\end{equation}
This is the leading degeneracy of the sine-quadrature search.

The numerical implication is direct. 
A target coefficient \(|\chioff|=10^{-3}\) corresponds to an equivalent frame error
\begin{equation}
|\dframe|
\simeq
\frac{10^{-3}}{3}
\simeq
3.33\times10^{-4}\ {\rm rad}
\simeq
0.019^\circ
\simeq
69'' .
\label{eq:intro_delta_scale_1e_minus_3}
\end{equation}
Likewise, \(|\chioff|=10^{-2}\) corresponds to \(0.19^\circ\), or about \(688''\). 
These are not exotic scales for phase, projection, reference-frame, ephemeris, or Earth-response residuals. 
A measured sine-quadrature coefficient therefore requires an explicit estimation model before it can be assigned a physical interpretation.

This framework deliberately separates three quantities that must not be identified with one another. 
The first is a dimensionless Halilsoy-inspired morphology ratio,
\begin{equation}
\etaH(\alpha,t,\rho)
=
3\frac{C_H(\alpha,t,\rho)}{\PH(t,\rho)} ,
\label{eq:intro_etaH}
\end{equation}
which describes a \emph{local} cross-to-plus tensorial structure in the weak-field tidal block of the Halilsoy sector. 
Away from zeros of \(\PH\), one may introduce the model-dependent local bookkeeping relation
\begin{equation}
\chioff^{\rm loc}(t)
=
\gamH\,\etaH(\alpha,t,\rho),
\label{eq:intro_chioff}
\end{equation}
where \(\gamH\) is an unknown phenomenological amplitude or coupling scale, together with the corresponding local acceleration conversion
\begin{equation}
\As^{\rm loc}(t)
=
a_0(t)\,\chioff^{\rm loc}(t).
\label{eq:intro_As}
\end{equation}
Equations~(\ref{eq:intro_chioff})--(\ref{eq:intro_As}) are not imposed as a global estimator identity in the forecast and are not continued through plus-sector zeros. 
They are local morphology-to-amplitude bookkeeping relations, not predictions of an absolute lunar acceleration generated by a matched Earth--Moon--Halilsoy spacetime. 
The operational forecast introduced in Section~\ref{sec:observation_model} instead fits the two finite, generic three-dimensional coefficients \(\chi_1\) and \(\chi_2\).

This distinction also resolves the apparent singular behavior of the ratio \(3C_H/\PH\). 
Near zeros of the plus-sector denominator \(\PH\), the ratio \(\etaH\) may become large even though the underlying finite tensor components need not diverge. 
The ratio is a morphology variable, not a globally valid amplitude coordinate. 
Accordingly, the finite quantities used through such sign changes are the tensor components themselves and the quadrant-preserving eigenframe angle, while the quantitative gravimetric forecast uses independently fitted finite coefficients \((\chi_1,\chi_2)\). 
The Halilsoy sector can therefore motivate a cross-type tensor morphology, but it cannot by itself normalize a lunar residual acceleration without an additional coupling or matching calculation.

The phenomenological tensor used to define the local off-diagonal channel is
\begin{equation}
\E_{\rm off}
=
\kM
\begin{pmatrix}
2 & \chioff
\\
\chioff & -1
\end{pmatrix}.
\label{eq:intro_off_tensor}
\end{equation}
Since this tensor is real and symmetric, its eigenvectors remain orthogonal whenever the eigenvalues are distinct. 
The off-diagonal coefficient rotates the entire eigenframe; it does not destroy the \(90^\circ\) separation of the principal axes. 
The rotation angle is
\begin{equation}
\Theta(\chioff)
=
\frac12
\arctan
\left(
\frac{2\chioff}{3}
\right)
=
\frac{\chioff}{3}
+
O(\chioff^3).
\label{eq:intro_theta_chioff}
\end{equation}
The eigenvalue shifts begin at order \(\chioff^2\). 
Accordingly, the residual interpretation is perturbative and should be restricted to \(|\chioff|\ll1\) unless a separate large-amplitude physical model is supplied.

A real gravimetric test cannot be based only on the idealized two-dimensional angular formula. 
For a station \(i\), a schematic residual gravity equation is
\begin{align}
\Delta g_i
&=
\hat{u}_i^a\,\Delta E_{ab}(t_i)\,r_i^b
+
\Delta g_i^{\rm solid}
+
\Delta g_i^{\rm ocean}
+
\Delta g_i^{\rm atm}
\nonumber
\\
&\quad
+
\Delta g_i^{\rm hyd}
+
\Delta g_i^{\rm polar}
+
\Delta g_i^{\rm inst}
+
\Delta g_i^{\rm site}
+
\epsilon_i .
\label{eq:intro_observation_model}
\end{align}
Here \(\hat{u}_i^a\) is the local projection direction, \(r_i^b\) denotes the displacement vector used in this schematic tensor-to-acceleration contraction, and \(\Delta E_{ab}\) is the residual tidal tensor after reference modeling. 
For the direct vertical tidal acceleration used in the quantitative forecast, Section~\ref{sec:observation_model} sets \(r_i^b=R_E\hat{u}_i^b\) explicitly. 
Elastic and self-gravitational Earth response are \emph{not} absorbed into an adjustable ``effective lever arm'' in that forecast; uncertainty in the terrestrial response is represented separately by station-specific response gain and lag nuisance parameters. 
The remaining terms represent imperfectly modeled solid-Earth tide, ocean loading, atmospheric loading, hydrological loading, polar motion, instrumental drift, site response, and noise.

For orientation only, a local two-dimensional linearized residual fit may be written as
\begin{equation}
y_i
=
\Ac X_{c,i}
+
\As X_{s,i}
+
\sum_{k=1}^{N_b} b_k Z_{k,i}
+
\epsilon_i,
\qquad
X_{c,i}=\cos(2\beta_i),
\qquad
X_{s,i}=\sin(2\beta_i),
\label{eq:intro_regression}
\end{equation}
where \(Z_{k,i}\) are nuisance templates. 
With covariance matrix \(C\), the corresponding generalized least-squares covariance is
\begin{equation}
{\rm Cov}(\hat{\bm\theta})
=
\left(
X^T C^{-1}X
\right)^{-1},
\qquad
\bm\theta
=
(\Ac,\As,b_1,\ldots,b_{N_b})^T .
\label{eq:intro_covariance}
\end{equation}
Within this \emph{local illustrative} parameterization one may write
\begin{equation}
\hat{\chioff}
=
\frac{\hat{\As}}{a_0},
\qquad
\sigma(\chioff)
=
\frac{\sigma(\As)}{a_0}.
\label{eq:intro_chioff_estimator}
\end{equation}
Equations~(\ref{eq:intro_regression})--(\ref{eq:intro_chioff_estimator}) are not the quantitative estimator used below. 
They only display the local sine-quadrature logic and the role of correlated nuisance structure. 
All numerical forecasts in Sections~\ref{sec:observation_model}--\ref{sec:sensitivity} use the globally defined three-dimensional two-component model with primary parameters \((\chi_1,\chi_2)\).

\textbf{The central contribution is a globally coherent residual-estimation framework that establishes an exact identifiability limit and quantifies how external information and model mismatch govern the attainable sensitivity.} The contribution of this paper is fivefold.
First, it formulates the cross-type lunar term as a phenomenological three-dimensional residual tensor rather than as a direct detection signature.
Second, it derives both the familiar two-dimensional frame leakage and its exact three-dimensional counterpart, showing that the leading lunar-direction error is exactly signal-degenerate at first order.
Third, it separates the Halilsoy-inspired morphology ratio from the fitted residual coefficients and does not use \(\gamma_H\) as the primary forecast parameter.
Fourth, it constructs a joint-prior GLS/Fisher model in which frame, ephemeris, projection, terrestrial-response, loading, polar-motion, colored-noise, and free geophysical nuisance terms are treated inside the inference problem rather than as post-hoc uncertainty floors.
Fifth, it subjects the forecast to matched and deliberately misspecified injection--recovery ensembles, including a genuinely held-out calibration/validation split for the Monte-Carlo-calibrated robust covariance.

The paper is organized as follows. Section~\ref{sec:newtonian_and_tensor} derives the Newtonian plus channel, the local off-diagonal basis, and the perturbative eigenframe interpretation. Section~\ref{sec:frame_degeneracy} derives the frame-misregistration degeneracy. Section~\ref{sec:halilsoy_template} defines the Halilsoy-inspired morphology ratio while separating it from the fitted residual amplitude. Section~\ref{sec:observation_model} develops the coherent three-dimensional station response and the joint-prior GLS/Fisher model. Section~\ref{sec:sensitivity} presents the injection--recovery robustness analysis and the formal-versus-robust sensitivity results. Section~\ref{sec:discussion} discusses interpretation and limitations, and Section~\ref{sec:conclusion} concludes.


\section{Newtonian tensor, sine basis, and eigenframe geometry}
\label{sec:newtonian_and_tensor}

The aim of this section is to define the local tensor channel used throughout the paper. 
The construction is intentionally local. 
It does not replace the three-dimensional theory of Earth tides, but it isolates the two angular quadratures that arise when the dominant Newtonian lunar tidal tensor is projected in a plane containing the modeled Earth--Moon direction.

\subsection{Newtonian lunar tidal tensor}

Let \(\bm D\) be the vector from the Earth's center to the Moon, let \(D=|\bm D|\), and define the lunar direction
\begin{equation}
\bm n_M
=
\frac{\bm D}{D}.
\label{eq:nM_def}
\end{equation}
For a local displacement \(\bm r\) from the Earth's center with \(|\bm r|\ll D\), the Newtonian lunar potential is
\begin{equation}
\Phi_M(\bm r)
=
-\frac{G\MM}{|\bm D-\bm r|}.
\label{eq:moon_potential}
\end{equation}
Expanding about the Earth's center and retaining the quadrupolar term gives
\begin{equation}
\Phi_{\rm tide}
=
-\frac{G\MM}{2D^3}
\left[
3(\bm n_M\cdot\bm r)^2-r^2
\right].
\label{eq:newtonian_tidal_potential}
\end{equation}
This is the standard leading-order Newtonian lunar tidal potential used in celestial mechanics and geophysical tide theory \citep{MurrayDermott1999,Melchior1983,Cartwright1999,Agnew2007,PoissonWill2014}.

In the local principal frame aligned with \(\bm n_M\), the three-dimensional Newtonian tidal tensor is
\begin{equation}
\E_N^{(3)}
=
\kM
\begin{pmatrix}
2 & 0 & 0
\\
0 & -1 & 0
\\
0 & 0 & -1
\end{pmatrix},
\qquad
\kM
=
\frac{G\MM}{D^3}.
\label{eq:newtonian_3d_tensor}
\end{equation}
It is trace-free:
\begin{equation}
\tr \E_N^{(3)}
=
2\kM-\kM-\kM
=
0.
\label{eq:newtonian_tracefree}
\end{equation}
The corresponding two-dimensional block in the plane spanned by the Earth--Moon direction and one transverse direction is
\begin{equation}
\E_N
=
\kM
\begin{pmatrix}
2 & 0
\\
0 & -1
\end{pmatrix}.
\label{eq:newtonian_2d_tensor}
\end{equation}
The eigenvalues are \(2\kM\) and \(-\kM\), and the eigenvectors are aligned with the Earth--Moon direction and the selected transverse direction. 
Thus the ordinary Newtonian lunar tide is a diagonal plus-type tensor in this frame.

The absence of an off-diagonal component in Eq.~(\ref{eq:newtonian_2d_tensor}) is the relevant point:
\begin{equation}
(\E_N)_{xy}=0.
\label{eq:newtonian_no_cross}
\end{equation}
An independent cross-type residual cannot be identified with a mere projection of this diagonal tensor. 
It must appear as an additional off-diagonal component or, equivalently, as a sine-quadrature term in the projected acceleration.

\subsection{Projected Newtonian plus channel}

Let
\begin{equation}
\bm n(\beta)
=
\begin{pmatrix}
\cos\beta
\\
\sin\beta
\end{pmatrix}
\label{eq:n_beta_newtonian}
\end{equation}
be a unit direction making an angle \(\beta\) with the modeled Earth--Moon principal axis. 
At the Earth's surface, write
\begin{equation}
\bm r(\beta)
=
\RE\bm n(\beta).
\label{eq:r_beta}
\end{equation}
The projected Newtonian tidal acceleration along this direction is
\begin{equation}
a_N(\beta)
=
\RE\,\bm n^T\E_N\bm n.
\label{eq:projected_newtonian_def}
\end{equation}
Substituting Eq.~(\ref{eq:newtonian_2d_tensor}) gives
\begin{align}
a_N(\beta)
&=
\kM\RE
\left(
2\cos^2\beta-\sin^2\beta
\right)
\nonumber
\\
&=
a_0
\left[
\frac12+\frac32\cos(2\beta)
\right],
\label{eq:newtonian_projected_acceleration}
\end{align}
where
\begin{equation}
a_0
=
\kM\RE
=
\frac{G\MM\RE}{D^3}.
\label{eq:a0_def_newtonian}
\end{equation}

Equation~(\ref{eq:newtonian_projected_acceleration}) contains a constant part and the plus-type harmonic \(\cos(2\beta)\). 
It contains no independent \(\sin(2\beta)\) term in the Newtonian principal frame. 
At \(\beta=45^\circ\),
\begin{equation}
a_N(45^\circ)
=
\frac12a_0.
\label{eq:newtonian_45_projection}
\end{equation}
This value is an ordinary Newtonian projection. 
It is not a cross-tidal residual.

\subsection{Phenomenological off-diagonal residual basis}

To define the residual basis, introduce the symmetric local extension
\begin{equation}
\E_{\rm off}
=
\kM
\begin{pmatrix}
2 & \chioff
\\
\chioff & -1
\end{pmatrix},
\label{eq:offdiag_tensor_main}
\end{equation}
where \(\chioff\) is a dimensionless projected residual coefficient. 
At this stage no physical source is assigned to \(\chioff\). 
It is a coefficient to be estimated or bounded from residual data.

The projected acceleration is
\begin{align}
a_{\rm off}(\beta)
&=
\RE\,\bm n^T\E_{\rm off}\bm n
\nonumber
\\
&=
a_0
\left[
2\cos^2\beta-\sin^2\beta
+
2\chioff\sin\beta\cos\beta
\right]
\nonumber
\\
&=
a_0
\left[
\frac12+\frac32\cos(2\beta)
+
\chioff\sin(2\beta)
\right].
\label{eq:projected_chioff_acceleration}
\end{align}
Thus the Newtonian plus part and the off-diagonal residual part occupy distinct quadrature channels:
\begin{align}
a_{\rm off}(\beta)
&=
a_N(\beta)+a_{\rm cross}(\beta),
\label{eq:harmonic_decomposition_chioff}
\\
a_N(\beta)
&=
a_0
\left[
\frac12+\frac32\cos(2\beta)
\right],
\label{eq:newtonian_plus_channel}
\\
a_{\rm cross}(\beta)
&=
a_0\chioff\sin(2\beta).
\label{eq:cross_residual_channel}
\end{align}

The residual term vanishes along the plus-aligned directions,
\begin{equation}
\beta
=
0^\circ,\ 90^\circ,\ 180^\circ,\ 270^\circ,
\label{eq:cross_zero_angles}
\end{equation}
and is extremal when
\begin{equation}
|\sin(2\beta)|=1.
\label{eq:cross_extrema_condition}
\end{equation}
Thus the extrema occur at
\begin{equation}
\beta
=
45^\circ,\ 135^\circ,\ 225^\circ,\ 315^\circ.
\label{eq:cross_extrema_angles}
\end{equation}
This is the precise sense in which the residual is \(45^\circ\)-type. 
It is not the statement that Newtonian gravity lacks a \(45^\circ\) projection. 
It is the statement that an independent off-diagonal residual has its extrema along the \(45^\circ\)-type directions.

\subsection{Eigenframe rotation and perturbative eigenvalues}

The same distinction appears at the level of the principal frame. 
For a real symmetric matrix
\begin{equation}
M
=
\begin{pmatrix}
a & b
\\
b & c
\end{pmatrix},
\label{eq:generic_symmetric_matrix}
\end{equation}
the rotation angle of the principal axes satisfies
\begin{equation}
\tan(2\Theta)
=
\frac{2b}{a-c}.
\label{eq:general_rotation_angle}
\end{equation}
For Eq.~(\ref{eq:offdiag_tensor_main}), \(a=2\), \(b=\chioff\), and \(c=-1\). 
Therefore
\begin{equation}
\Theta(\chioff)
=
\frac12
\arctan
\left(
\frac{2\chioff}{3}
\right).
\label{eq:theta_chioff}
\end{equation}
For \(|\chioff|\ll1\),
\begin{equation}
\Theta(\chioff)
=
\frac{\chioff}{3}
+
O(\chioff^3).
\label{eq:theta_chioff_small}
\end{equation}
The eigenvectors remain orthogonal because the tensor is symmetric. 
The off-diagonal residual rotates the entire eigenframe; it does not make the two principal axes non-orthogonal.

\begin{figure}[t]
\centering
\includegraphics[width=0.98\linewidth]{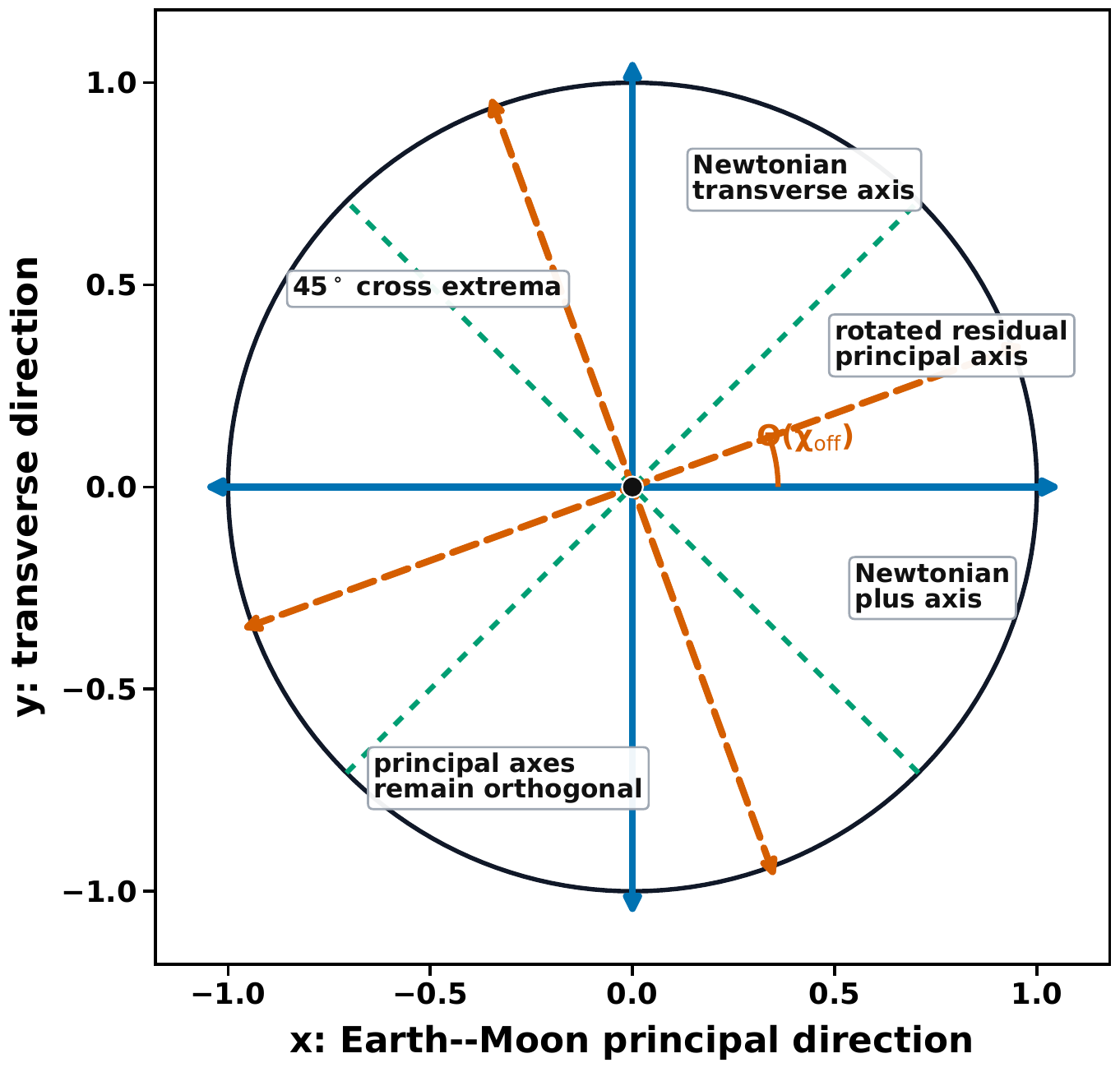}
\caption{
\textbf{Local Newtonian principal frame and rotated off-diagonal residual frame.}
The blue axes represent the ordinary Newtonian principal frame aligned with the Earth--Moon direction and one selected transverse direction. 
A symmetric local off-diagonal coefficient \(\chioff\) rotates the whole eigenframe by \(\Theta(\chioff)\) while preserving the \(90^\circ\) separation of the two principal axes. 
The dotted \(45^\circ\)-type directions indicate where the local sine-quadrature basis is extremal. 
This is a two-dimensional geometric illustration only: it is neither the global three-dimensional forecast geometry of Section~\ref{sec:observation_model} nor an absolute amplitude prediction.
}
\label{fig:eigenframe_schematic}
\end{figure}

Figure~\ref{fig:eigenframe_schematic} summarizes this local eigenframe rotation and distinguishes the \(45^\circ\)-type extrema of the sine quadrature from the globally projected three-dimensional signal used later.

The eigenvalues of Eq.~(\ref{eq:offdiag_tensor_main}) are
\begin{equation}
\lambda_\pm
=
\frac{\kM}{2}
\left[
1
\pm
\sqrt{9+4\chioff^2}
\right].
\label{eq:offdiag_eigenvalues_exact}
\end{equation}
At \(\chioff=0\), these reduce to
\begin{equation}
\lambda_+=2\kM,
\qquad
\lambda_-=-\kM.
\label{eq:offdiag_eigenvalues_newtonian_limit}
\end{equation}
For small \(|\chioff|\),
\begin{equation}
\sqrt{9+4\chioff^2}
=
3
+
\frac{2}{3}\chioff^2
+
O(\chioff^4),
\label{eq:sqrt_expansion_chioff}
\end{equation}
and hence
\begin{align}
\lambda_+
&=
2\kM
+
\frac{\kM}{3}\chioff^2
+
O(\chioff^4),
\label{eq:lambda_plus_small_chioff}
\\
\lambda_-
&=
-\kM
-
\frac{\kM}{3}\chioff^2
+
O(\chioff^4).
\label{eq:lambda_minus_small_chioff}
\end{align}
The eigenframe rotation is first order in \(\chioff\), whereas the eigenvalue changes are second order. 
This justifies the perturbative interpretation only when
\begin{equation}
|\chioff|\ll1.
\label{eq:small_chioff_regime}
\end{equation}
Large values of \(|\chioff|\) can be studied algebraically, but they should not be interpreted as small lunar residuals without an additional physical model.

This section defines the local tensor channel. 
A possible off-diagonal lunar residual is not identified by the Newtonian value at \(45^\circ\), but by a sine-quadrature residual coefficient and a corresponding perturbative eigenframe rotation. 
The next section shows why this same sine-quadrature channel is strongly degenerate with ordinary frame and projection errors.


\section{Frame-misregistration degeneracy of the sine-quadrature channel}
\label{sec:frame_degeneracy}

Section~\ref{sec:newtonian_and_tensor} identified the sine-quadrature channel as the natural tensorial location of a symmetric off-diagonal residual. 
However, this channel is not uniquely diagnostic of a physical off-diagonal lunar tide. 
A small error in the modeled Newtonian principal frame converts part of the ordinary diagonal plus channel into an apparent sine-quadrature contribution. 
This is the dominant interpretive degeneracy of the proposed residual search.

Let the true Newtonian principal-frame angle be \(\beta-\dframe\), while the residual analysis is performed using the nominal angle \(\beta\). 
The purely diagonal Newtonian projection then becomes
\begin{equation}
a_N(\beta-\dframe)
=
a_0
\left[
\frac12
+
\frac32
\cos\bigl(2\beta-2\dframe\bigr)
\right].
\label{eq:frame_shift_start}
\end{equation}
Using
\begin{equation}
\cos(2\beta-2\dframe)
=
\cos(2\beta)\cos(2\dframe)
+
\sin(2\beta)\sin(2\dframe),
\end{equation}
one obtains
\begin{equation}
a_N(\beta-\dframe)
=
a_0
\left[
\frac12
+
\frac32\cos(2\beta)\cos(2\dframe)
+
\frac32\sin(2\beta)\sin(2\dframe)
\right].
\label{eq:frame_shift_expanded}
\end{equation}
Thus a purely diagonal Newtonian tide, when expressed in a slightly misregistered frame, contains an apparent \(\sin(2\beta)\) component.

The apparent sine-quadrature amplitude generated by frame error alone is
\begin{equation}
\Asframe
=
\frac32 a_0\sin(2\dframe).
\label{eq:As_frame}
\end{equation}
The corresponding dimensionless apparent coefficient is
\begin{equation}
\chi_{\rm app}
=
\frac{\Asframe}{a_0}
=
\frac32\sin(2\dframe).
\label{eq:chiapp_exact}
\end{equation}
For \(|\dframe|\ll1\),
\begin{equation}
\chi_{\rm app}
\simeq
3\dframe.
\label{eq:chiapp_small_angle}
\end{equation}
This relation is independent of any Halilsoy-inspired morphology. 
It is a purely Newtonian projection effect. 
The relation \(\chi_{\rm app}\simeq3\dframe\) is the local two-dimensional manifestation of the exact three-dimensional design-column proportionality derived in Section~\ref{sec:observation_model}, where each lunar tangent-direction derivative is exactly three times the corresponding cross-residual signal column.

\begin{figure*}[t]
\centering
\includegraphics[width=0.98\textwidth]{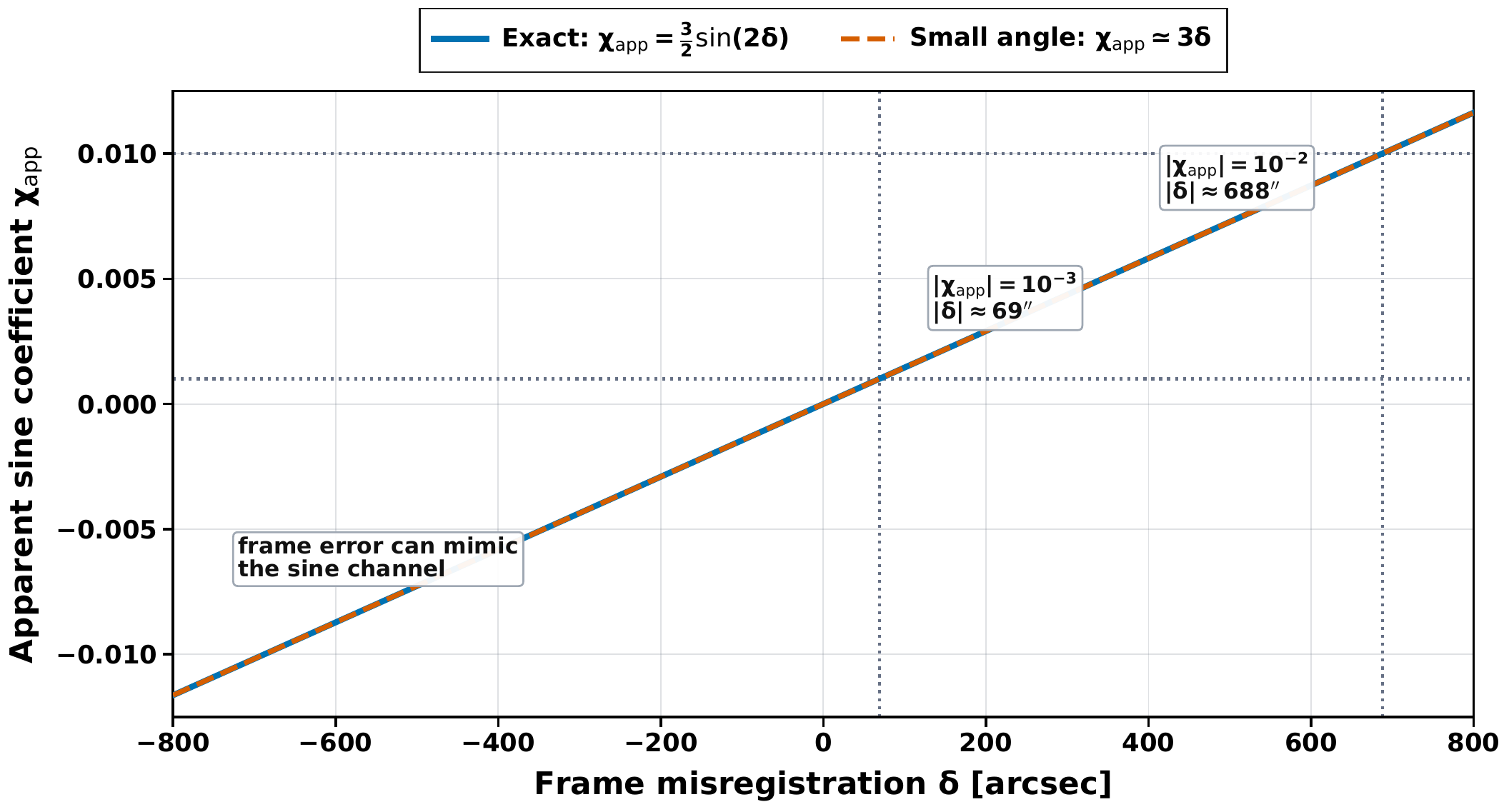}
\caption{
\textbf{Local analytic illustration of frame-misregistration degeneracy.}
A purely diagonal Newtonian tide written in a slightly misregistered local two-dimensional frame, \(\beta\rightarrow\beta-\dframe\), produces an apparent sine coefficient \(\chi_{\rm app}=(3/2)\sin(2\dframe)\simeq3\dframe\). The solid curve is the exact expression and the dashed curve is its small-angle approximation. 
The coefficient-to-angle benchmark is given in Eq.~(\ref{eq:intro_delta_scale_1e_minus_3}) and collected with the acceleration conversion in Figure~\ref{fig:sensitivity_budget}. 
The quantitative forecast does not use this scalar-angle approximation as an external error floor; Section~\ref{sec:observation_model} derives the exact three-dimensional signal--frame column proportionality and includes the angular directions jointly with external priors.
}
\label{fig:frame_degeneracy}
\end{figure*}

\begin{figure*}[t]
\centering
\includegraphics[width=0.92\textwidth]{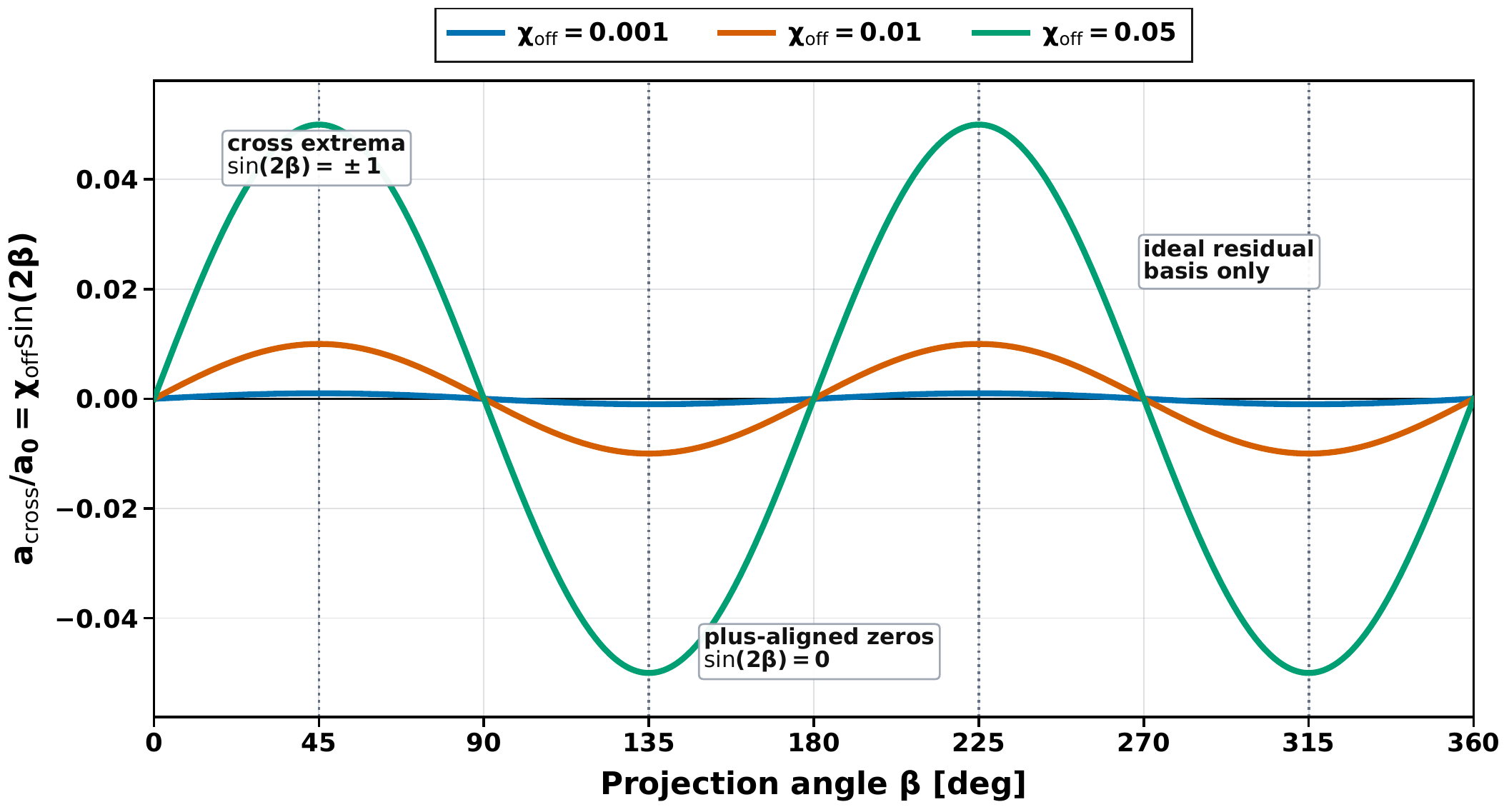}
\caption{
\textbf{Local illustrative residual-only sine-quadrature basis.}
After removal of the modeled Newtonian plus contribution in the local two-dimensional construction, the phenomenological coefficient \(\chioff\) enters as \(a_{\rm cross}(\beta)/a_0=\chioff\sin(2\beta)\); the three curves show the declared values \(\chioff=10^{-3},10^{-2},\) and \(5\times10^{-2}\). 
The extrema occur along the \(45^\circ\)-type directions and the zeros along the plus-aligned directions. 
This figure visualizes the local analytic basis retained for interpretation; it is not the global design matrix used in the quantitative forecast, which is built from the three-dimensional tensor and station contraction in Section~\ref{sec:observation_model}.
}
\label{fig:cross_residual_only}
\end{figure*}

Figures~\ref{fig:frame_degeneracy} and \ref{fig:cross_residual_only} respectively show the frame-induced apparent coefficient and the local residual-only sine quadrature. 
The exact expression \(\chi_{\rm app}=(3/2)\sin(2\dframe)\) and the small-angle approximation \(\chi_{\rm app}\simeq3\dframe\) are nearly indistinguishable over the displayed range. 
Consequently, a fitted sine coefficient at the level \(|\chioff|\sim10^{-3}\)--\(10^{-2}\) cannot be interpreted as a physical off-diagonal residual unless frame, projection, ephemeris, local-vertical, station-coordinate, and terrestrial-response uncertainties are propagated at comparable levels.

The benchmark conversions in Eqs.~(\ref{eq:intro_delta_scale_1e_minus_3}) and (\ref{eq:chiapp_small_angle}) set the relevant angular scale; Figure~\ref{fig:frame_degeneracy} displays the same relation without treating it as a measured frame error or an external error floor.

A fitted sine coefficient should therefore be interpreted schematically as
\begin{equation}
\chi_{\rm fit}
=
\chioff
+
\chi_{\rm app}
+
\chi_{\rm load}
+
\chi_{\rm atm}
+
\chi_{\rm hyd}
+
\chi_{\rm polar}
+
\chi_{\rm inst}
+
\chi_{\rm site}
+
\cdots .
\label{eq:chi_fit_decomposition}
\end{equation}
Here \(\chioff\) is the projected physical off-diagonal residual coefficient, while the remaining terms denote apparent sine-quadrature contributions from frame, loading, atmospheric, hydrological, polar-motion, instrumental, and local site effects. 
Only the component that remains after nuisance modeling and covariance marginalization can be associated with a candidate physical off-diagonal residual.

This point changes the interpretation of the sine-quadrature channel. 
An off-diagonal residual must contribute to the \(\sin(2\beta)\) basis in the local two-dimensional model, but the converse is not true: a measured sine coefficient does not by itself prove an off-diagonal lunar tide. 
In that local illustration, the relevant quantity is a fitted or bounded residual amplitude \(\As\) interpreted jointly with nuisance templates and their covariance. 
The operational three-dimensional forecast does not fit a single global \(\As\); it fits the two-component coefficient vector \(\bchi=(\chi_1,\chi_2)^T\).

In acceleration units, the frame-induced sine signal is
\begin{equation}
|\Asframe|
\simeq
3a_0|\dframe|.
\label{eq:Asframe_small_angle}
\end{equation}
Using \(a_0\simeq55\,\mu{\rm Gal}\), a frame misregistration of \(69''\) gives an apparent amplitude of approximately \(0.055\,\mu{\rm Gal}\), while a misregistration of \(688''\) gives approximately \(0.55\,\mu{\rm Gal}\). 
These are the same acceleration scales associated with \(|\chioff|=10^{-3}\) and \(|\chioff|=10^{-2}\), respectively. 
Thus acceleration sensitivity and frame control are inseparable.

\begin{figure*}[t]
\centering
\includegraphics[width=0.92\textwidth]{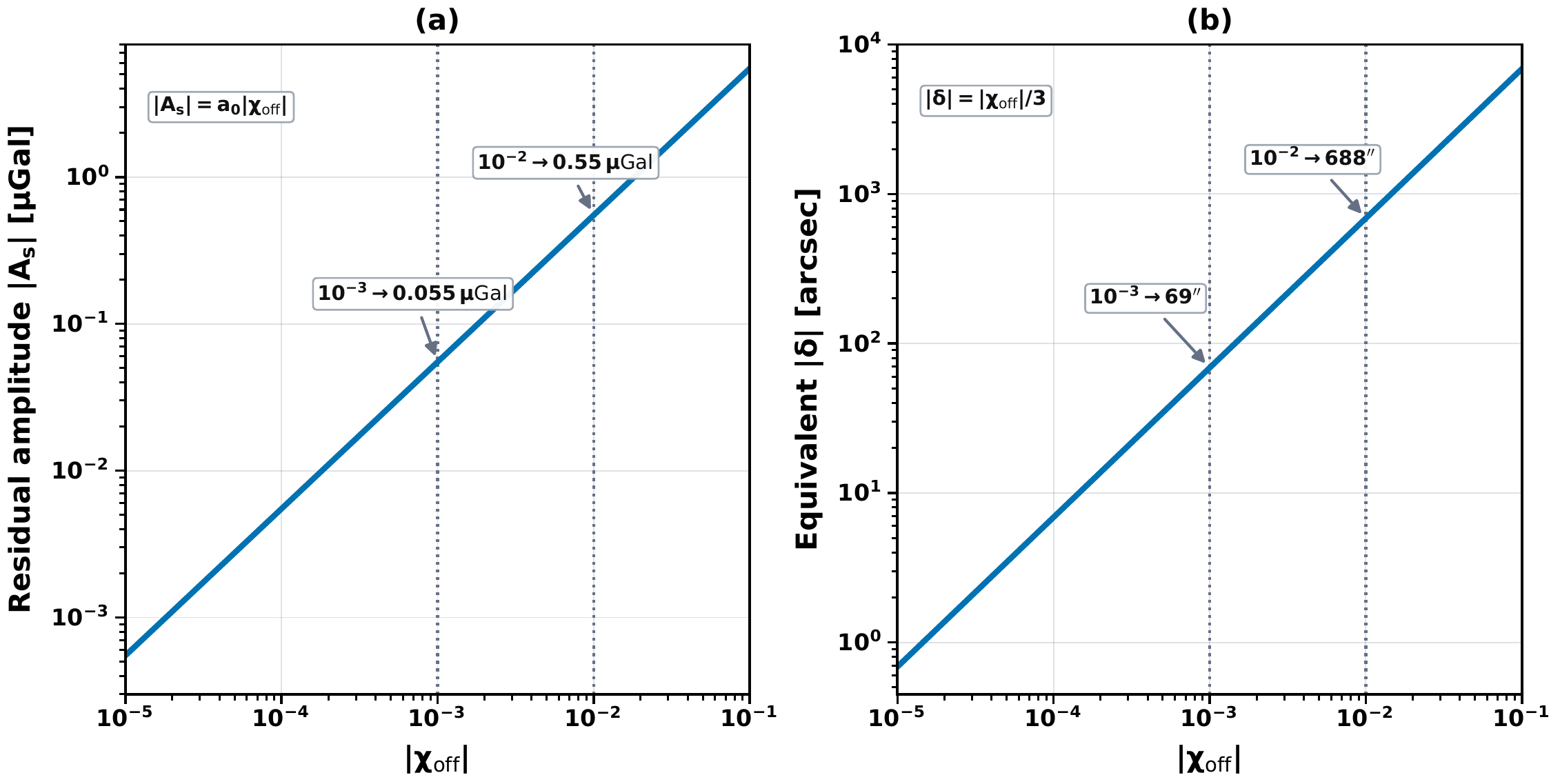}
\caption{
\textbf{Local benchmark conversion between residual coefficient, acceleration scale, and frame-control scale.}
(a) The left panel converts \(|\chioff|\) into the residual acceleration scale \(|A_s|=a_0|\chioff|\), using the representative value \(a_0\simeq55\,\mu{\rm Gal}\) only as an intuitive benchmark. (b) The right panel converts the same coefficient into the equivalent frame-control scale using the local small-angle relation \(|\dframe|\simeq|\chioff|/3\). 
For example, \(|\chioff|=10^{-3}\) corresponds to about \(0.055\,\mu{\rm Gal}\) and \(69''\), while \(|\chioff|=10^{-2}\) corresponds to about \(0.55\,\mu{\rm Gal}\) and \(688''\). 
The angular axis is therefore an equivalent frame-misregistration scale, not a measured frame error, confidence limit, or universally allowable tolerance.
These are scale comparisons rather than statistical forecast outputs. 
The quantitative analysis in Sections~\ref{sec:observation_model}--\ref{sec:sensitivity} uses the time-dependent lunar scale \(a_0(t)\), the full three-dimensional station response, and explicit prior-limited nuisance parameters.
}
\label{fig:sensitivity_budget}
\end{figure*}

Figure~\ref{fig:sensitivity_budget} collects the coefficient, acceleration, and equivalent-misregistration conversions in one place; it is a scale comparison rather than an additional statistical constraint.

Accordingly, the quantities \(\As\) and \(\chioff\) used in Sections~\ref{sec:newtonian_and_tensor}--\ref{sec:halilsoy_template} should be read as local illustrative residual coordinates, not as direct predictions of an absolute acceleration generated by the Halilsoy spacetime and not as the global forecast parameters. 
The quantitative estimator introduced in Section~\ref{sec:observation_model} uses the finite three-dimensional coefficients \((\chi_1,\chi_2)\).


\section{Halilsoy-inspired morphology template}
\label{sec:halilsoy_template}

Sections~\ref{sec:newtonian_and_tensor} and \ref{sec:frame_degeneracy} defined the local off-diagonal residual coefficient \(\chioff\) without assigning it a physical source. 
We now introduce the Halilsoy cross-polarized cylindrical-wave sector only as a morphology template for such an off-diagonal tidal block. 
This restricted interpretation is essential. 
The Halilsoy solution is not used here as a global Earth--Moon spacetime, and it does not by itself predict an absolute lunar residual acceleration.

For reference, the underlying cross-polarized cylindrical-wave geometry may be written in the Jordan--Ehlers--Kundt--Kompaneets form \citep{Halilsoy1988,GriffithsPodolsky2009}
\begin{equation}
ds^2
=
e^{2(\gamma-\psi)}
\left(
-dt^2+d\rho^2
\right)
+
\rho^2 e^{-2\psi}d\phi^2
+
e^{2\psi}
\left(
dz+\omega d\phi
\right)^2 .
\label{eq:halilsoy_metric}
\end{equation}
In the Halilsoy solution the functions \(\psi\), \(\omega\), and \(\gamma\) are determined by the wave amplitude, the length scale, and the cross-polarization parameter. 
In one common parametrization,
\begin{align}
e^{-2\psi(t,\rho)}
&=
e^{A J_0(\rho/\lambda)\cos(t/\lambda)}
\sinh^2\left(\frac{\alpha}{2}\right)
+
e^{-A J_0(\rho/\lambda)\cos(t/\lambda)}
\cosh^2\left(\frac{\alpha}{2}\right),
\label{eq:halilsoy_psi_function}
\\
\omega(t,\rho)
&=
-A\sinh\alpha\,
\rho J_1(\rho/\lambda)\sin(t/\lambda).
\label{eq:halilsoy_omega_function}
\end{align}
Here \(A\) is the wave amplitude, \(\lambda\) is the wave scale, and \(\alpha\) controls the second, cross-polarized sector. 
The function \(\gamma(t,\rho)\) is fixed by the remaining Einstein equations. 
The present paper does not use this metric as a global Earth--Moon spacetime. 
It uses only its local weak-field tidal morphology, in particular the fact that the cross-polarized sector generates an off-diagonal tidal block.

Halilsoy's cross-polarized cylindrical gravitational-wave solution extends the Einstein--Rosen class by introducing a second polarization through an off-diagonal metric sector \citep{EinsteinRosen1937,Halilsoy1988}. 
Cylindrical gravitational waves and their exact-solution generalizations provide a useful relativistic setting in which local tidal eigenframes can be rotated by cross-polarized structure \citep{WeberWheeler1957,Thorne1965,Stephani2003,GriffithsPodolsky2009}. 
Recent work on standing and cylindrical gravitational waves has further clarified the geometric and physical interpretation of such wave sectors \citep{SzybkaNaqvi2021,GlodSikoraSzybka2022,NikielSzybka2025,SzybkaZychowiczHunik2025}.

In general relativity, the local tidal field is encoded in the electric part of the curvature tensor. 
For neighboring freely falling test particles, the geodesic-deviation equation is
\begin{equation}
\ddot{\xi}^{\,i}
=
- E^i{}_{j}\xi^j,
\qquad
E_{ij}
=
R_{i0j0},
\label{eq:geodesic_deviation_halilsoy}
\end{equation}
where the components are evaluated in a local orthonormal frame \citep{Pirani1956,Misner1973,Wald1984,PoissonWill2014}. 
In weak-field transverse-traceless language, the tidal matrix is equivalently proportional to the second time derivative of the metric perturbation,
\begin{equation}
E_{ij}
=
-\frac12
\partial_t^2 h^{\rm TT}_{ij},
\label{eq:tt_tidal_relation}
\end{equation}
up to sign conventions \citep{Maggiore2007}. 
Thus plus and cross gravitational-wave polarizations have direct local tidal meaning.

\subsection{Weak-field transverse tidal block}

Consider the weak-field transverse sector of the Halilsoy geometry in a local cylindrical orthonormal frame
\begin{equation}
(\bm e_\rho,\bm e_\phi,\bm e_z).
\label{eq:cylindrical_frame_halilsoy}
\end{equation}
The second-polarization sector acts in the transverse \((\bm e_\phi,\bm e_z)\) plane. 
In that plane, the weak-field tidal block may be written as
\begin{equation}
\E_\perp^H
=
\frac{\eps}{2\lambda^2}
\begin{pmatrix}
-Q_0+\dfrac{\lambda}{\rho}Q_1
&
-\sinh\alpha\,W
\\
-\sinh\alpha\,W
&
Q_0
\end{pmatrix}.
\label{eq:halilsoy_tidal_block}
\end{equation}
Here \(\eps\) is a weak-field amplitude, \(\lambda\) is the wave scale, and \(\alpha\) controls the second-polarization sector.

The functions appearing in Eq.~(\ref{eq:halilsoy_tidal_block}) are
\begin{align}
Q_0(t,\rho)
&=
J_0(\rho/\lambda)\cos(t/\lambda),
\label{eq:Q0_def}
\\
Q_1(t,\rho)
&=
J_1(\rho/\lambda)\cos(t/\lambda),
\label{eq:Q1_def}
\\
W(t,\rho)
&=
J_1(\rho/\lambda)\sin(t/\lambda).
\label{eq:W_def}
\end{align}
The variables \(t/\lambda\) and \(\rho/\lambda\) are dimensionless. 
Equivalently, units are chosen so that the characteristic wave scale \(\lambda\) nondimensionalizes both time and radius; in relativistic units this corresponds to setting \(c=1\).

The off-diagonal component of the Halilsoy tidal block is
\begin{equation}
E^H_{\phi z}
=
-\frac{\eps}{2\lambda^2}
\sinh\alpha\,W(t,\rho).
\label{eq:halilsoy_off_component}
\end{equation}
It vanishes when \(\alpha=0\), and it is generally nonzero for \(\alpha\neq0\). 
Thus the parameter \(\alpha\) controls the cross-polarized tidal sector in this local weak-field representation.

The diagonal contrast of the transverse block is
\begin{align}
E^H_{\phi\phi}-E^H_{zz}
&=
\frac{\eps}{2\lambda^2}
\left(
-Q_0+\frac{\lambda}{\rho}Q_1-Q_0
\right)
\nonumber
\\
&=
-\frac{\eps}{2\lambda^2}
\left(
2Q_0-\frac{\lambda}{\rho}Q_1
\right).
\label{eq:halilsoy_plus_contrast}
\end{align}
It is useful to define the plus-sector factor
\begin{equation}
\PH(t,\rho)
=
2Q_0(t,\rho)-\frac{\lambda}{\rho}Q_1(t,\rho).
\label{eq:PH_def}
\end{equation}
Substituting Eqs.~(\ref{eq:Q0_def}) and (\ref{eq:Q1_def}) gives
\begin{equation}
\PH(t,\rho)
=
\left[
2J_0(\rho/\lambda)
-
\frac{\lambda}{\rho}J_1(\rho/\lambda)
\right]
\cos(t/\lambda).
\label{eq:PH_explicit}
\end{equation}

For compactness, define
\begin{equation}
\DHal(\rho)
=
2J_0(\rho/\lambda)
-
\frac{\lambda}{\rho}J_1(\rho/\lambda),
\label{eq:DH_def}
\end{equation}
so that
\begin{equation}
\PH(t,\rho)
=
\DHal(\rho)\cos(t/\lambda).
\label{eq:PH_DH_form}
\end{equation}
Similarly, define the cross-sector numerator
\begin{equation}
C_H(\alpha,t,\rho)
=
\sinh\alpha\,W(t,\rho)
=
\sinh\alpha\,J_1(\rho/\lambda)\sin(t/\lambda).
\label{eq:CH_def}
\end{equation}
The pair \(C_H\) and \(\PH\) describes the local cross-to-plus morphology of the weak-field Halilsoy tidal block.

\subsection{Eigenframe ratio}

Because Eq.~(\ref{eq:halilsoy_tidal_block}) is a real symmetric \(2\times2\) matrix, its principal axes remain orthogonal wherever the eigenvalues are distinct. 
The off-diagonal term rotates the eigenframe in the \((\bm e_\phi,\bm e_z)\) plane. 
Using the standard relation
\begin{equation}
\tan(2\Theta)
=
\frac{2b}{a-c},
\label{eq:halilsoy_generic_rotation}
\end{equation}
one obtains for the Halilsoy block
\begin{equation}
\tan(2\Theta_H)
=
\frac{2C_H(\alpha,t,\rho)}{\PH(t,\rho)}.
\label{eq:halilsoy_rotation_ratio}
\end{equation}
In explicit Bessel form,
\begin{equation}
\tan(2\Theta_H)
=
\frac{
2\sinh\alpha\,J_1(\rho/\lambda)\sin(t/\lambda)
}{
\left[
2J_0(\rho/\lambda)
-
\dfrac{\lambda}{\rho}J_1(\rho/\lambda)
\right]
\cos(t/\lambda)
}.
\label{eq:halilsoy_rotation_ratio_explicit}
\end{equation}

Near sign changes of the plus-sector denominator, it is preferable to preserve quadrant information by using a two-argument arctangent. 
Define
\begin{equation}
\NH(\alpha,t,\rho)
=
2\sinh\alpha\,J_1(\rho/\lambda)\sin(t/\lambda).
\label{eq:NH_def}
\end{equation}
Then the local Halilsoy eigenframe angle can be written as
\begin{equation}
\Theta_H(\alpha,t,\rho)
=
\frac12
\atanTwo
\left(
\NH(\alpha,t,\rho),
\DHal(\rho)\cos(t/\lambda)
\right).
\label{eq:ThetaH_atan2}
\end{equation}
This angle remains finite through sign changes of the plus-sector factor, even when the ratio \(C_H/\PH\) becomes a poor variable.

\subsection{Morphology ratio and local illustrative amplitude mapping}

To compare the local Halilsoy cross-to-plus morphology with the phenomenological lunar residual tensor, define the dimensionless morphology ratio
\begin{equation}
\etaH(\alpha,t,\rho)
=
3\frac{C_H(\alpha,t,\rho)}{\PH(t,\rho)}.
\label{eq:etaH_def}
\end{equation}
The factor \(3\) is the diagonal contrast \(2-(-1)\) of the local Newtonian two-dimensional block. 
Substitution gives
\begin{equation}
\etaH(\alpha,t,\rho)
=
3
\frac{
\sinh\alpha\,J_1(\rho/\lambda)\sin(t/\lambda)
}{
\DHal(\rho)\cos(t/\lambda)
}.
\label{eq:etaH_explicit}
\end{equation}
Where \(\cos(t/\lambda)\neq0\) and \(\DHal(\rho)\neq0\), this may be written as
\begin{equation}
\etaH(\alpha,t,\rho)
=
3\sinh\alpha
\frac{J_1(\rho/\lambda)}{\DHal(\rho)}
\tan(t/\lambda).
\label{eq:etaH_tangent_form}
\end{equation}

The quantity \(\etaH\) is not an absolute lunar acceleration coefficient. 
It is a dimensionless morphology ratio of the local Halilsoy tidal block. 
The absolute weak-field curvature scale cancels from this ratio. 
Therefore multiplying \(\etaH\) directly by the lunar scale \(a_0\) would assume, rather than derive, a coupling between the cylindrical-wave morphology and the localized Earth--Moon field.

For local comparison \emph{away from plus-sector zeros}, one may introduce the illustrative mapping
\begin{equation}
\chioff^{\rm loc}(t)
=
\gamH\,\etaH(\alpha,t,\rho),
\label{eq:chioff_epsH_etaH}
\end{equation}
where \(\gamH\) is an unknown phenomenological coupling or amplitude scale. 
The corresponding local acceleration conversion is
\begin{equation}
\As^{\rm loc}(t)
=
a_0(t)\chioff^{\rm loc}(t)
=
a_0(t)\gamH\etaH(\alpha,t,\rho).
\label{eq:As_epsH_etaH}
\end{equation}
These equations are bookkeeping relations for the local morphology only. 
They are not predictions from a matched Earth--Moon--Halilsoy spacetime, they are not continued through \(\PH=0\), and they are not imposed on the global forecast parameters \(\chi_1,\chi_2\).

\subsection{Behavior near plus-sector zeros}

The separation between \(\etaH\), \(\chioff\), and \(\As\) is especially important near zeros of the plus-sector denominator. 
When \(\PH(t,\rho)\) approaches zero while \(C_H(\alpha,t,\rho)\) remains finite, the ratio \(\etaH=3C_H/\PH\) can become large. 
This does not imply a divergent lunar acceleration. 
It means only that a cross-to-plus ratio is not a good amplitude variable near a plus-sector zero.

The finite geometric description is instead carried by the underlying tensor components and by the quadrant-preserving eigenframe angle \(\Theta_H\) in Eq.~(\ref{eq:ThetaH_atan2}). 
In particular, if \(\gamH\) were held fixed and nonzero, the local ratio-based quantity \(\gamH\etaH\) would inherit the divergence of \(\etaH\) as \(\PH\to0\); its finiteness is therefore \emph{not} automatic. 
For that reason the relation \(\chioff^{\rm loc}=\gamH\etaH\) is used only away from plus-sector zeros and is not adopted as a global signal template. 
The quantitative gravimetric estimator is formulated directly in terms of the finite three-dimensional tensor coefficients \(\chi_1,\chi_2\), while \(\Theta_H\) is evaluated with \(\atanTwo\) when the local Halilsoy eigenframe is discussed.

\subsection{Small-\texorpdfstring{\(\alpha\)}{alpha} limit}

For small \(\alpha\),
\begin{equation}
\sinh\alpha
=
\alpha
+
O(\alpha^3).
\label{eq:sinh_alpha_small}
\end{equation}
Thus
\begin{equation}
\etaH(\alpha,t,\rho)
=
3\alpha
\frac{J_1(\rho/\lambda)}{\DHal(\rho)}
\tan(t/\lambda)
+
O(\alpha^3),
\label{eq:etaH_small_alpha}
\end{equation}
away from plus-sector zeros. 
The corresponding local illustrative mapping is
\begin{equation}
\chioff^{\rm loc}(t)
=
3\gamH\alpha
\frac{J_1(\rho/\lambda)}{\DHal(\rho)}
\tan(t/\lambda)
+
O(\gamH\alpha^3),
\label{eq:chioff_small_alpha}
\end{equation}
again only away from plus-sector zeros. 
The Halilsoy-inspired morphology is linear in the second-polarization parameter at leading order, but no global terrestrial amplitude follows without a separate physical normalization and a finite matched signal construction.

The role of the Halilsoy sector in the present construction is therefore morphological rather than dynamical. 
It supplies an explicit relativistic example of a cross-polarized off-diagonal tidal block. 
The operational gravimetric problem in the next section is deliberately formulated independently of the ratio \(\eta_H\) and independently of \(\gamma_H\), using the finite generic coefficients \((\chi_1,\chi_2)\).


\section{Residual gravimetric observation model}
\label{sec:observation_model}

The preceding two-dimensional construction identifies the local sine-quadrature channel, but a gravimetric test requires one station-independent three-dimensional tensor. The lunar direction generally has components along both local horizontal axes, so assigning every station its own vertical--meridional signal plane does not define a coherent global residual. We therefore retain the local algebra as an interpretive guide and construct the numerical forecast directly from a three-dimensional station response.

This construction is not a substitute for an operational Earth-tide code. Standard lunar--solar forcing and Earth-orientation conventions remain the reference model \citep{Doodson1921,CartwrightEdden1973,Tamura1987,HartmannWenzel1995,PetitLuzum2010,Agnew2007}. The question addressed here is narrower: how would one additional phenomenological cross-residual tensor project into the station data stream, and how would uncertainty in reference geometry, terrestrial response, loading, environment, polar motion, calibration, drift, and network covariance propagate into its estimate? The resulting model is an intermediate step between the local ansatz and a station-specific real-data reduction, designed to expose identifiability and robustness limits without treating synthetic inputs as observatory measurements.

\subsection{Three-dimensional residual tensor and station response}

Let \(\hat{\bm n}_M(t)\) be the instantaneous unit vector from the Earth toward the Moon. At each time choose two orthonormal transverse vectors \(\bm q^{(1)}(t)\) and \(\bm q^{(2)}(t)\) satisfying
\begin{equation}
\hat{\bm n}_M\cdot\bm q^{(A)}=0,
\qquad
\bm q^{(A)}\cdot\bm q^{(B)}=\delta_{AB}.
\label{eq:transverse_basis_orthonormal}
\end{equation}
For reproducibility, the numerical implementation constructs \(\bm q^{(1)}\) by projecting the terrestrial spin axis onto the plane orthogonal to \(\hat{\bm n}_M\), with a fixed fallback axis in the nongeneric near-parallel case, and defines
\(\bm q^{(2)}=\hat{\bm n}_M\times\bm q^{(1)}\).
A rotation of \(\bm q^{(1)},\bm q^{(2)}\) within the transverse plane changes only the polarization coordinates; because both coefficients are fitted, the two-dimensional signal subspace is invariant.

The phenomenological lunar cross-residual tensor is
\begin{equation}
\Delta E^{\rm cross}_{ab}(t)
=
\kM(t)
\sum_{A=1}^{2}
\chi_A
\left[
n_{M,a}q^{(A)}_b
+
q^{(A)}_a n_{M,b}
\right],
\label{eq:cross_tensor_3d}
\end{equation}
where
\begin{equation}
\kM(t)=\frac{GM_M}{D_M^3(t)},
\qquad
a_0(t)=\kM(t)R_E .
\label{eq:time_dependent_lunar_scale}
\end{equation}
Equation~(\ref{eq:cross_tensor_3d}) is symmetric and trace free. The primary forecast parameters are therefore
\begin{equation}
\bchi=(\chi_1,\chi_2)^T.
\label{eq:primary_chi_vector}
\end{equation}
They are generic, dimensionless phenomenological residual coefficients. They are not identified with \(\gamma_H\) in the forecast.

The choice of two coefficients is not an arbitrary duplication of the earlier local sine-quadrature parameter. At a given lunar direction the plane perpendicular to \(\hat{\bm n}_M\) is two dimensional. Unless an independently derived physical model selects a unique transverse polarization, a globally defined residual tensor must therefore allow both independent transverse components. A rotation of the pair \(\{\bm q^{(1)},\bm q^{(2)}\}\) simply rotates the coordinate vector \((\chi_1,\chi_2)\) and leaves the physical two-dimensional signal subspace unchanged. This basis covariance is useful operationally: no station is assigned its own preferred cross direction, and all stations are fitted to one global tensor model.

The time dependence in the signal columns arises from geometry and the lunar scale \(a_0(t)\), not from a fitted time-dependent \(\chi_A\). In the present methodological forecast \(\chi_1\) and \(\chi_2\) are constant coefficients multiplying a known time-dependent design. This is intentionally different from fitting \(\gamma_H\). A \(\gamma_H\) estimator would require a specified physical map from the Halilsoy morphology to the Earth--Moon system and would carry the additional factor \(\eta_H(\alpha,t,\rho)\) through the complete three-dimensional projection. Keeping these two statistical targets separate avoids using the local morphology ratio as though it were a globally regular amplitude coordinate.

For a gravimeter whose local vertical is \(\hat{\bm u}_i(t)\), the direct vertical response of the cross sector is
\begin{align}
\Delta g^{\rm cross}_i(t)
&=
R_E\,
u_i^a
\Delta E^{\rm cross}_{ab}(t)
u_i^b
\nonumber\\
&=
2a_0(t)
\sum_{A=1}^{2}
\chi_A
\bigl(\hat{\bm u}_i\cdot\hat{\bm n}_M\bigr)
\bigl(\hat{\bm u}_i\cdot\bm q^{(A)}\bigr).
\label{eq:station_cross_response_3d}
\end{align}
Thus the two signal columns are
\begin{equation}
S_{A,i}(t)
=
2a_0(t)
\bigl(\hat{\bm u}_i\cdot\hat{\bm n}_M\bigr)
\bigl(\hat{\bm u}_i\cdot\bm q^{(A)}\bigr),
\qquad
A=1,2.
\label{eq:signal_columns_3d}
\end{equation}
The lunar scale \(a_0(t)\) is evaluated with the time-dependent lunar distance; the frequently quoted value \(a_0\simeq55\,\mu{\rm Gal}\) is used only as a benchmark scale and not as a constant forecast normalization.

The corresponding direct Newtonian lunar vertical tide is
\begin{equation}
g_{N,i}^{M}(t)
=
a_0(t)
\left[
3\bigl(\hat{\bm u}_i\cdot\hat{\bm n}_M\bigr)^2-1
\right].
\label{eq:newtonian_vertical_3d}
\end{equation}
The solar reference tide is represented analogously, with the solar direction and the corresponding time-dependent solar scale. The solar term is included in the projection and nuisance derivatives below, so the forecast is not a lunar-only residual separation.

\subsection{Linearized frame leakage and exact first-order degeneracy}

The dominant identifiability limitation is easiest to see directly in three dimensions. Perturb the modeled lunar direction within its tangent plane,
\begin{equation}
\hat{\bm n}_M
\longrightarrow
\hat{\bm n}_M
+
\sum_{A=1}^{2}
\delta_A\bm q^{(A)}
+
O(\delta^2).
\label{eq:lunar_direction_perturbation}
\end{equation}
Differentiating Eq.~(\ref{eq:newtonian_vertical_3d}) gives
\begin{equation}
\left.
\frac{\partial g_N^M}{\partial\delta_A}
\right|_{\delta=0}
=
6a_0(t)
\bigl(\hat{\bm u}\cdot\hat{\bm n}_M\bigr)
\bigl(\hat{\bm u}\cdot\bm q^{(A)}\bigr).
\label{eq:3d_frame_derivative}
\end{equation}
From Eq.~(\ref{eq:station_cross_response_3d}),
\begin{equation}
\frac{\partial \Delta g^{\rm cross}}{\partial\chi_A}
=
2a_0(t)
\bigl(\hat{\bm u}\cdot\hat{\bm n}_M\bigr)
\bigl(\hat{\bm u}\cdot\bm q^{(A)}\bigr).
\label{eq:3d_signal_derivative}
\end{equation}
Therefore
\begin{equation}
\boxed{
\left.
\frac{\partial g_N^M}{\partial\delta_A}
\right|_{\delta=0}
=
3
\frac{\partial \Delta g^{\rm cross}}{\partial\chi_A}
}
\qquad (A=1,2).
\label{eq:exact_3d_frame_degeneracy}
\end{equation}

\begin{proposition}[Exact first-order signal--frame degeneracy]
\label{prop:exact_degeneracy}
For the Newtonian vertical response of Eq.~(\ref{eq:newtonian_vertical_3d}) and the two-component cross-residual response of Eq.~(\ref{eq:station_cross_response_3d}), the first-order derivative columns satisfy
\(\partial g_N^M/\partial\delta_A=3\,\partial\Delta g^{\rm cross}/\partial\chi_A\)
for every station, every epoch, and both transverse components \(A=1,2\).
Consequently, a data-only Fisher matrix containing \((\chi_1,\chi_2)\) together with the lunar-direction parameters \((\delta_1,\delta_2)\) is exactly singular in the corresponding two directions, and any finite posterior uncertainty on \(\chi_A\) is inherited from external angular information rather than from the gravimetric data alone.
\end{proposition}

This exact proportionality is the identifiability constraint carried into the joint model below.

\subsection{Station-level projection and terrestrial-response nuisance parameters}

The local vertical is not treated as exact. With local north and east tangent vectors
\(\hat{\bm e}_{N,i}\) and \(\hat{\bm e}_{E,i}\), write
\begin{equation}
\hat{\bm u}_i
\longrightarrow
\hat{\bm u}_i
+
\epsilon_{N,i}\hat{\bm e}_{N,i}
+
\epsilon_{E,i}\hat{\bm e}_{E,i}
+
O(\epsilon^2).
\label{eq:local_vertical_perturbation}
\end{equation}
For the direct lunar plus solar reference tide, the corresponding first-order projection derivatives are
\begin{align}
\frac{\partial g_i}{\partial\epsilon_{N,i}}
&=
6a_M
(\hat{\bm u}_i\cdot\hat{\bm n}_M)
(\hat{\bm e}_{N,i}\cdot\hat{\bm n}_M)
+
6a_S
(\hat{\bm u}_i\cdot\hat{\bm n}_S)
(\hat{\bm e}_{N,i}\cdot\hat{\bm n}_S),
\label{eq:projection_north_derivative}\\
\frac{\partial g_i}{\partial\epsilon_{E,i}}
&=
6a_M
(\hat{\bm u}_i\cdot\hat{\bm n}_M)
(\hat{\bm e}_{E,i}\cdot\hat{\bm n}_M)
+
6a_S
(\hat{\bm u}_i\cdot\hat{\bm n}_S)
(\hat{\bm e}_{E,i}\cdot\hat{\bm n}_S).
\label{eq:projection_east_derivative}
\end{align}
The numerical model also includes two lunar-frame tangent parameters, two lunar-ephemeris tangent parameters, two solar-ephemeris tangent parameters, station-specific north/east projection parameters, station-specific terrestrial-response gain and lag parameters, and two polar-motion amplitudes. Loading is not represented by a single network-wide proxy. For every station, the fitted prior-controlled block contains independent sine and cosine ocean-loading quadratures for M2, N2, O1, S2, K2, and P1, equivalent to a station- and constituent-specific complex amplitude response, together with a separate local long-period loading term. Atmospheric and hydrological transfer residuals are also represented by distinct station-specific columns: the atmospheric proxy uses synthetic periods \(2.7\), \(5.3\), and \(9.1\) days, whereas the hydrological proxy uses \(29.5\), \(61\), and \(120\) days. These are declared synthetic transfer bases rather than station products inferred from observations.
The response-gain and response-lag columns are phenomenological first-order representations of uncertainty in the effective elastic/self-gravitational Earth response, including local Love-number/response-amplitude and phase uncertainty; they are not interpreted as changes in the direct geometric lever arm \(R_E\). 
Lunar frame and lunar ephemeris columns have the same leading likelihood direction in this synthetic model; their distinct external priors represent distinct contributions to the total direction uncertainty, and the data are not claimed to separate them.

In an operational reduction, these reference columns would be generated from a conventional lunar--solar tide model rather than from the compact ephemeris surrogate used here. The tensor form specifies the instantaneous geometric projection, while the harmonic representation organizes the forcing into constituent frequencies and phases \citep{Doodson1921,CartwrightEdden1973,Tamura1987,HartmannWenzel1995}; the Stress tests below connect the two descriptions by injecting constituent-level lunar and solar mismatch.

\subsection{Joint-prior GLS/Fisher sensitivity forecast}

Generalized least squares (GLS) extends ordinary least squares to data with
correlated and/or heteroscedastic errors by weighting the residuals with the
inverse covariance matrix \citep{Aitken1935}. For a Gaussian linear model,
the corresponding Fisher information matrix quantifies the local information
carried by the data about the fitted parameters and is given by
\(F_{\rm data}=J^{T}C^{-1}J\); when this data-only Fisher matrix is nonsingular,
its inverse gives the parameter covariance \citep{Fisher1922,Cramer1946}.

The stacked residual model is
\begin{equation}
\bm y
=
S\bchi
+
J_{\rm ang}\bm\delta
+
Z\bm\alpha
+
\bm\epsilon,
\label{eq:joint_linear_model_3d}
\end{equation}
where \(S=(S_1,S_2)\) contains the two cross-residual columns, \(J_{\rm ang}\) contains frame, ephemeris, and projection derivatives, and \(Z\) contains the remaining terrestrial-response and geophysical nuisance columns. The noise covariance is \(C\).

Gaussian external information on the controlled nuisance parameters is represented by a prior precision matrix \(P\). The posterior Fisher matrix is
\begin{equation}
F_{\rm post}
=
J^TC^{-1}J+P,
\label{eq:posterior_fisher}
\end{equation}
where \(J=[S,J_{\rm ang},Z]\). The corresponding joint-prior GLS estimator is
\begin{equation}
\hat{\bm\theta}_{\rm GLS}
=
F_{\rm post}^{-1}
\left(
J^TC^{-1}\bm y
+
P\bm\theta_{\rm prior}
\right),
\label{eq:joint_prior_gls_estimator}
\end{equation}
where \(\bm\theta_{\rm prior}\) contains the external prior means. In the numerical forecast the controlled nuisance parameters are centered on their reference values, so their centered prior means vanish. The GLS estimator in Eq.~(\ref{eq:joint_prior_gls_estimator}) is the inference engine used throughout the forecast and in every injection--recovery realization. The Fisher matrix supplies its formal local sensitivity: the joint covariance is
\begin{equation}
{\rm Cov}(\hat{\bm\theta})
=
F_{\rm post}^{-1}.
\label{eq:posterior_covariance_joint}
\end{equation}
The Monte Carlo analysis below does not replace this GLS/Fisher calculation. It tests whether the formal covariance remains calibrated when repeated synthetic data are generated under either the fitted model or the declared mismatch budget.
For numerical conditioning diagnostics we form the dimensionless normalized Fisher matrix
\begin{equation}
\widetilde F
=
D^{-1/2}F_{\rm post}D^{-1/2},
\qquad
D=\operatorname{diag}(F_{\rm post}),
\label{eq:normalized_fisher}
\end{equation}
so the reported condition number is invariant to simple rescaling of individual parameter units. 
The covariance is transformed back to the physical parameter units after inversion; a Moore--Penrose pseudoinverse is reserved for an actual rank deficiency of the normalized Fisher matrix at the stated numerical tolerance.

The free geophysical block contains station offsets and linear drifts, station-specific diurnal and semidiurnal sine/cosine terms, and common fortnightly and monthly sine/cosine terms. The prior-controlled block contains the angular, ephemeris, projection, response, ocean/local-loading, atmospheric, hydrological, and polar-motion parameters described above. This replaces the earlier post-hoc quadrature-floor construction: the dominant frame/projection/ephemeris uncertainties and the declared loading/environmental transfer uncertainties are now inside the joint inference problem.

Within each station the baseline stochastic covariance is represented by a site-dependent irregular-sampling AR(1) precision model. For a time separation \(\Delta t\),
\begin{equation}
\phi(\Delta t)
=
\rho^{\Delta t/\Delta t_0},
\label{eq:irregular_ar1_phi}
\end{equation}
so data gaps are handled without pretending that separated samples are adjacent.

The block-diagonal site covariance \(C_0\) is augmented by a low-rank network term,
\begin{equation}
C=C_0+U\Lambda U^T,
\label{eq:network_covariance_model}
\end{equation}
where \(U\) combines sine/cosine temporal modes at \(2.5\), \(7\), \(20\), and \(60\) days with eigenmodes of the station correlation matrix. For stations \(i\) and \(j\), separated by angular distance \(d_{ij}\), the fitted spatial correlation is
\begin{equation}
R_{ij}=\delta_{ij}+(1-\delta_{ij})\rho_{\rm net}
\exp\!\left(-\frac{d_{ij}}{\ell_{\rm net}}\right).
\label{eq:network_spatial_correlation}
\end{equation}
The matrices are checked for positive definiteness, and \(C^{-1}\) is applied through the Woodbury identity rather than by forming the dense time-domain covariance. The Stress generator increases the network amplitude and correlation strength, shortens the angular correlation length, and shifts the temporal periods; it therefore tests covariance misspecification rather than merely redrawing from the fitted covariance.

\subsection{Covariance-weighted nuisance-subspace diagnostics}

Pairwise correlations alone do not measure the overlap between a two-dimensional signal subspace and a multidimensional nuisance space. Let \(Z_N\) denote a chosen nuisance block. The covariance-weighted projection of the signal space into that nuisance space is summarized by the eigenvalues of
\begin{equation}
\mathcal{R}
=
F_{SS}^{-1/2}
F_{SN}
F_{NN}^{+}
F_{NS}
F_{SS}^{-1/2},
\label{eq:canonical_r2_matrix}
\end{equation}
where the blocks are computed from the data Fisher matrix \(J^TC^{-1}J\), and \(+\) denotes the pseudoinverse. We quote the largest eigenvalue as \(R^2\). When the angular-control columns are included, Eq.~(\ref{eq:exact_3d_frame_degeneracy}) requires
\begin{equation}
R^2_{\rm all}=1
\label{eq:r2_exact_expected}
\end{equation}
up to numerical precision. We also report \(R^2_{\rm geo}\), computed using only the free geophysical nuisance block.

\subsection{Synthetic forecast scenarios}

The three scenarios are deliberately synthetic bounding configurations rather than representations of named stations. Their network sizes, durations, cadences, mean post-correction residual-noise levels, AR(1) coefficients, and principal angular priors are summarized in Table~\ref{tab:v13_assumptions}. A fixed random-gap fraction of \(3\%\) and one 12-hour outage per station are included in each scenario. Because network size, observing duration, cadence, residual-noise scale, and prior widths change together, the scenario-to-scenario improvement is not a controlled estimate of the gain from station number alone. Nor is it a performance advantage predicted by the Halilsoy morphology; it is the combined consequence of the declared synthetic observing configurations and external priors.

\begin{table*}[t]
\centering
\caption{\textbf{Synthetic joint-prior forecast assumptions.}
The station coordinates are synthetic. Angular widths are one-standard-deviation external priors. The noise levels are effective post-correction residual scales, not empirical estimates for named superconducting-gravimeter stations.}
\label{tab:v13_assumptions}
\footnotesize
\begin{tabular}{lcccccccc}
\hline
Scenario & \(N_{\rm sta}\) & Duration & Cadence & \(\sigma_n\) & \(\rho\) &
\(\sigma_{\rm frame}\) & \(\sigma_{\rm proj}\) & \(\sigma_{\rm eph}\)\\
 &  & (d) & (h) & (\(\mu{\rm Gal}\)) &  & (arcsec) & (arcsec) & (arcsec)\\
\hline
Conservative & 1  & 30  & 1.0 & 1.00 & 0.30 & 11.0 & 16.0 & 5.0\\
Moderate     & 3  & 90  & 1.0 & 0.30 & 0.20 &  6.0 &  8.0 & 1.0\\
Optimistic   & 10 & 180 & 0.5 & 0.10 & 0.10 &  1.5 &  2.5 & 0.7\\
\hline
\end{tabular}
\end{table*}

The synthetic station sets are
\begin{align}
\mathcal{S}_{\rm cons}
&=
\{(45^\circ,0^\circ)\},
\nonumber\\
\mathcal{S}_{\rm mod}
&=
\{(45^\circ,0^\circ),(0^\circ,120^\circ),(-45^\circ,240^\circ)\},
\nonumber\\
\mathcal{S}_{\rm opt}
&=
\{
(-60^\circ,0^\circ),(-45^\circ,45^\circ),(-30^\circ,90^\circ),
(-15^\circ,135^\circ),(0^\circ,180^\circ),
\nonumber\\
&\qquad
(15^\circ,225^\circ),(30^\circ,270^\circ),(45^\circ,315^\circ),
(60^\circ,60^\circ),(30^\circ,150^\circ)
\}.
\label{eq:v13_station_sets}
\end{align}
The response-gain priors are \(0.5\%\), \(0.2\%\), and \(0.1\%\); the corresponding lag priors are \(60\), \(20\), and \(5\) s. For the Conservative, Moderate, and Optimistic scenarios, respectively, the total station-specific ocean-loading prior scales are \(0.009\), \(0.0033\), and \(0.0017\,\mu{\rm Gal}\), and the local long-period loading priors use the same three scales. The separate atmospheric/hydrological prior scales are \((0.006,0.006)\), \((0.0022,0.0022)\), and \((0.0011,0.0011)\,\mu{\rm Gal}\). The fitted network-stochastic RMS values are \(0.020\), \(0.0070\), and \(0.0025\,\mu{\rm Gal}\), with correlation strengths \(0.30\), \(0.35\), and \(0.40\) and angular correlation lengths \(70^\circ\), \(60^\circ\), and \(50^\circ\). Polar-motion amplitude priors are \(0.005\), \(0.002\), and \(0.001\,\mu{\rm Gal}\).

\begin{figure*}[t]
\centering
\includegraphics[width=0.98\textwidth]{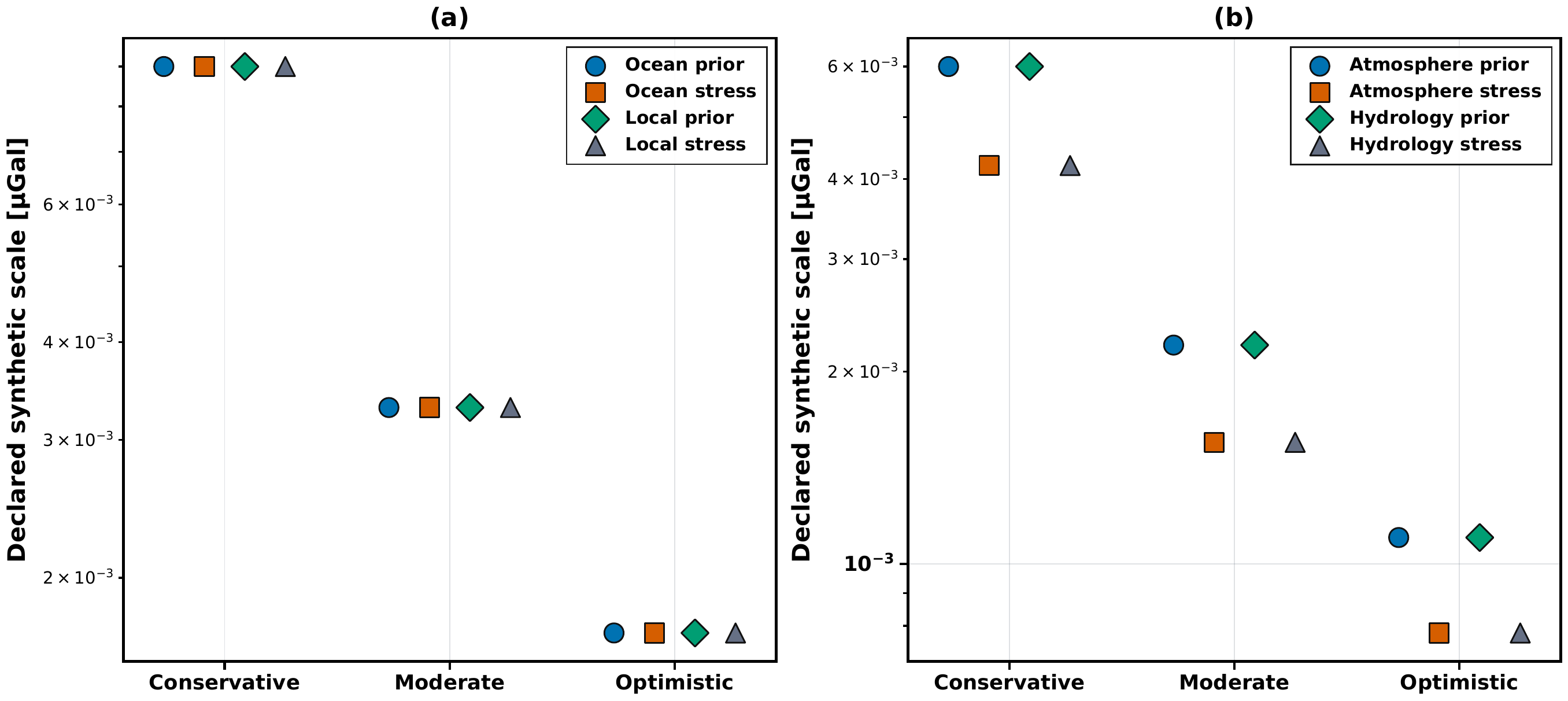}
\caption{\textbf{Declared station-loading and environmental-transfer scales.}
(a) The left panel compares the total station-specific ocean-loading and local long-period loading scales in the fitted-prior and Stress constructions. (b) The right panel gives the corresponding distinct atmospheric and hydrological transfer scales. Prior markers denote one-standard-deviation coefficient widths; Stress markers denote RMS amplitudes of omitted injections. A smaller Stress marker does not imply a milder matched perturbation because the Stress generator also changes temporal periods, station-dependent phases, and the stochastic network covariance. The atmospheric and hydrological terms use different temporal bases and remain separate processes throughout the calculation. These declared synthetic scales document model inputs rather than fit outcomes or observatory estimates.}
\label{fig:v13_loading_environment}
\end{figure*}

Figure~\ref{fig:v13_loading_environment} collects the fitted-prior and Stress scales; the caption distinguishes their definitions and the additional forms of mismatch carried by the Stress generator.

\subsection{Illustrative joint-prior forecast results}

Table~\ref{tab:v13_fisher_results} gives the marginalized two-component forecast. The quoted \(\sigma_{\rm worst}\) is the square root of the largest eigenvalue of the \(2\times2\) posterior covariance of \((\chi_1,\chi_2)\). The normalized joint-Fisher condition numbers are approximately \(118\), \(51\), and \(64\) for the Conservative, Moderate, and Optimistic scenarios, with \(39\), \(89\), and \(264\) fitted parameters, respectively.

\begin{table*}[t]
\centering
\caption{\textbf{Three-dimensional joint-prior GLS/Fisher forecast.}
Angular, ephemeris, projection, terrestrial-response, station/constituent ocean-loading, local-loading, atmospheric, hydrological, and polar-motion uncertainties are marginalized inside the Fisher matrix, with the cross-station stochastic covariance included in the GLS weighting. \(R^2_{\rm all}\) is the data-only canonical signal--nuisance overlap including the exact leading frame degeneracy, whereas \(R^2_{\rm geo}\) uses only the free geophysical nuisance block.}
\label{tab:v13_fisher_results}
\begin{tabular}{lcccccc}
\hline
Scenario & \(\sigma(\chi_1)\) & \(\sigma(\chi_2)\) & \(\sigma_{\rm worst}\) &
\(R^2_{\rm all}\) & \(R^2_{\rm geo}\) & Retained fraction\\
\hline
Conservative & \(1.567\times10^{-3}\) & \(9.512\times10^{-3}\) & \(9.513\times10^{-3}\) & 1.000 & 0.195 & 0.956\\
Moderate     & \(1.742\times10^{-4}\) & \(4.490\times10^{-4}\) & \(4.490\times10^{-4}\) & 1.000 & 0.091 & 0.963\\
Optimistic   & \(2.736\times10^{-5}\) & \(4.182\times10^{-5}\) & \(4.182\times10^{-5}\) & 1.000 & 0.073 & 0.968\\
\hline
\end{tabular}
\end{table*}

\begin{figure*}[t]
\centering
\includegraphics[width=0.90\textwidth]{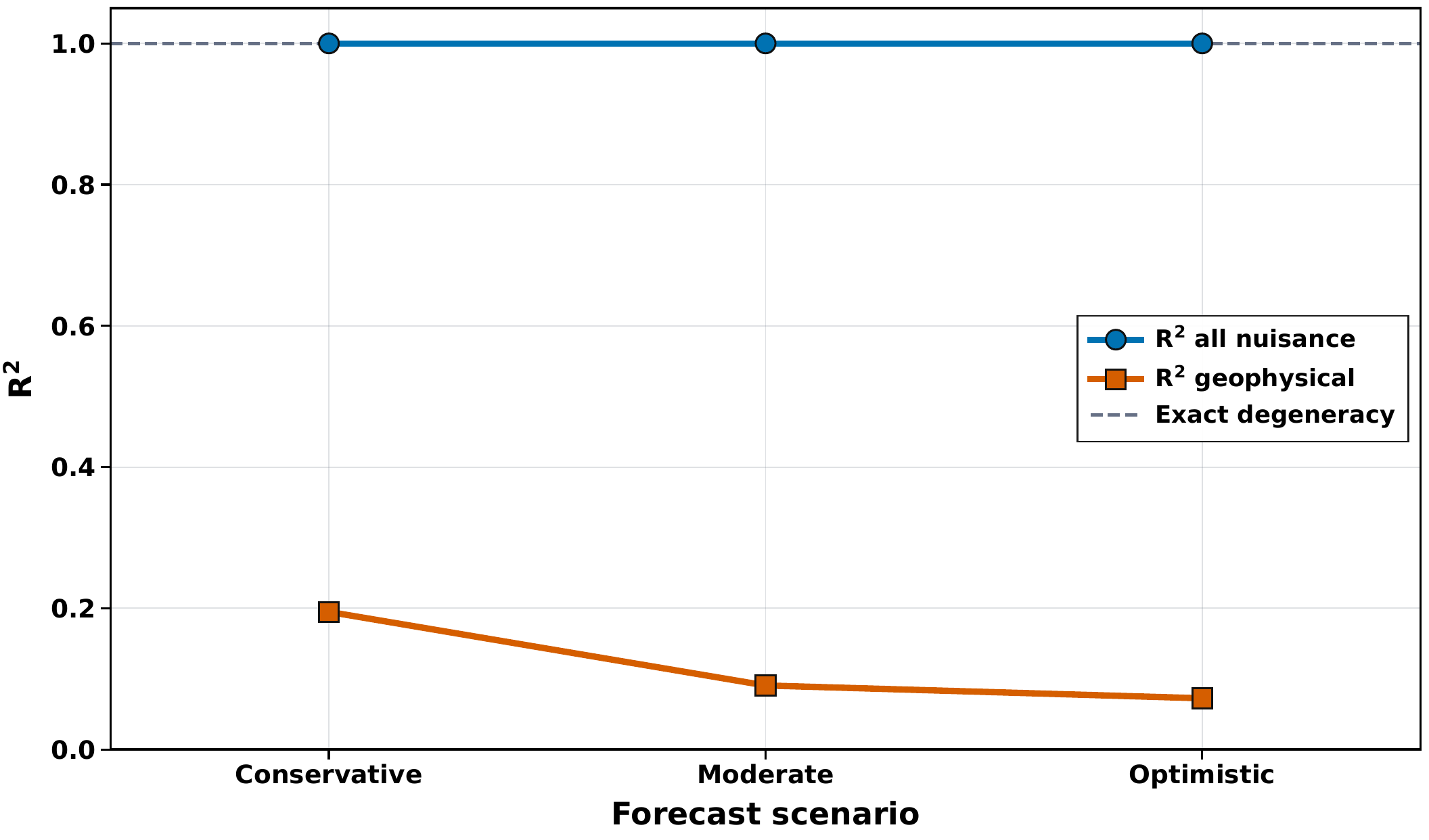}
\caption{\textbf{Covariance-weighted signal--nuisance subspace overlap.}
The blue curve gives \(R^2_{\rm all}\), including the leading lunar-frame/ephemeris control directions, and equals unity in every scenario because the corresponding first-order columns are exactly proportional to the two cross-residual signal columns. The orange curve gives \(R^2_{\rm geo}\) for the free geophysical nuisance block alone and decreases from 0.195 to 0.091 to 0.073 across the three declared configurations. The dashed line at unity marks the analytically expected exact degeneracy.}
\label{fig:v13_r2}
\end{figure*}

The exact value \(R^2_{\rm all}=1\) is the numerical manifestation of Eq.~(\ref{eq:exact_3d_frame_degeneracy}), not a failure of the forecast. Figure~\ref{fig:v13_r2} displays this exact overlap together with the smaller geophysical-only overlap. Across the declared configurations, \(R^2_{\rm geo}\) decreases while the controlled frame direction remains intrinsically prior limited.

\begin{figure*}[t]
\centering
\includegraphics[width=0.90\textwidth]{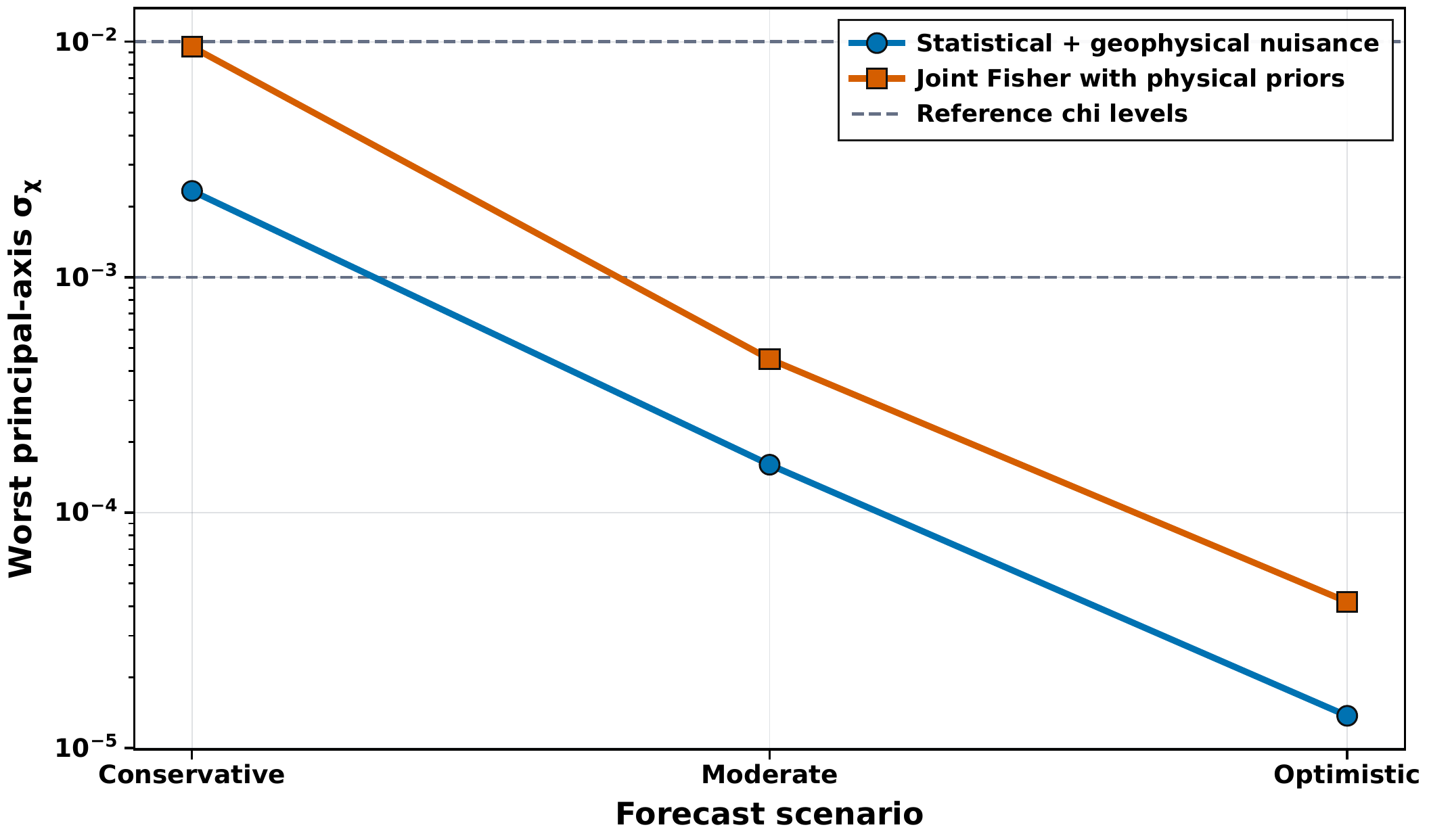}
\caption{\textbf{Formal statistical/geophysical sensitivity and joint-prior uncertainty.}
The blue curve shows the worst principal-axis uncertainty obtained after marginalizing over the free geophysical nuisance block, while the orange curve shows the full joint-prior posterior uncertainty. The increase from blue to orange is the quantitative cost of the prior-limited frame, ephemeris, projection, terrestrial-response, loading, and polar-motion uncertainties. Dashed horizontal lines mark reference coefficient scales \(10^{-3}\) and \(10^{-2}\). These are conditional forecast uncertainties, not measured residuals.}
\label{fig:v13_joint_uncertainty}
\end{figure*}

The widening gap between the two curves toward the high-sensitivity configurations motivates the prior-width sweep below.

\begin{figure*}[t]
\centering
\includegraphics[width=0.98\textwidth]{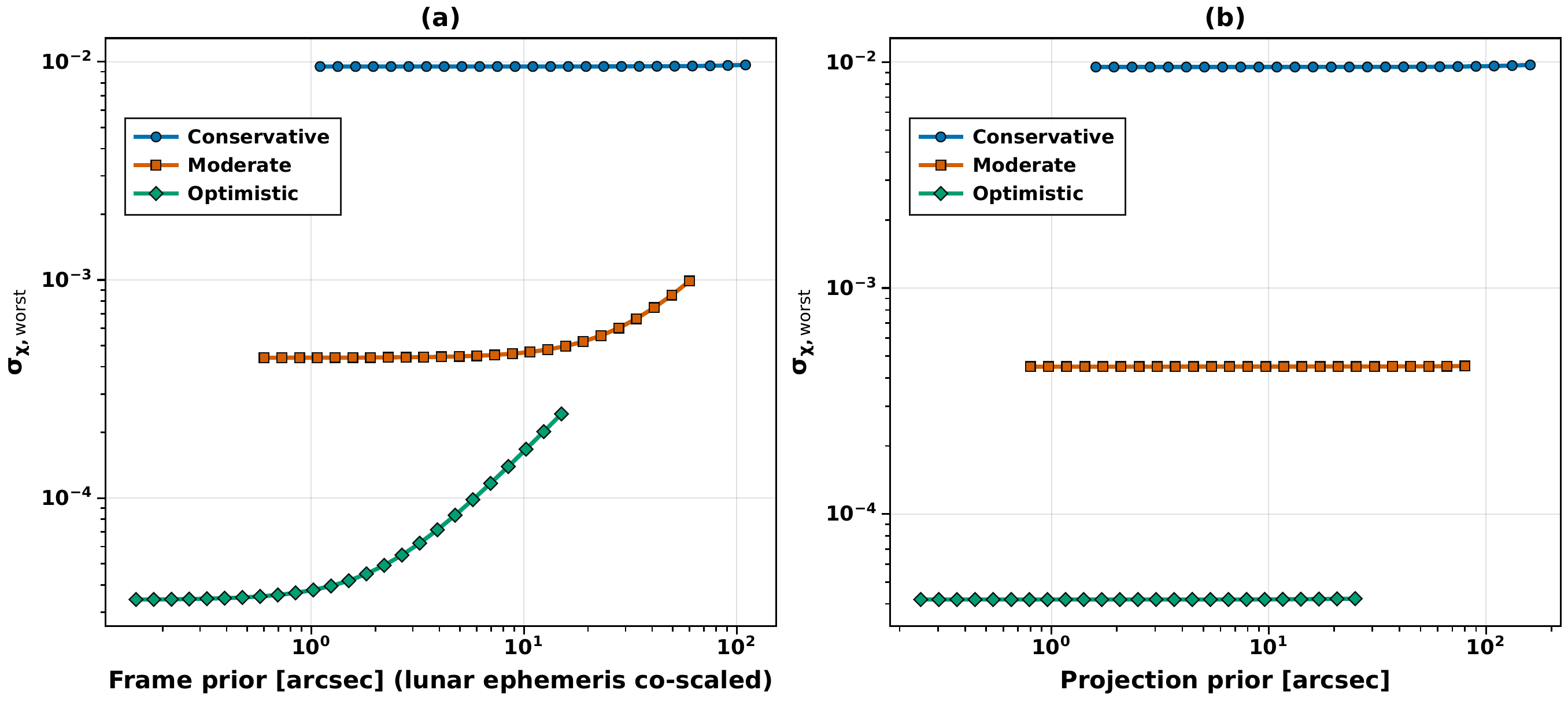}
\caption{\textbf{Prior-width dependence of the joint forecast.}
(a) The left panel shows the worst principal-axis uncertainty as the global lunar-frame and lunar-ephemeris priors are scaled together. (b) The right panel shows the same quantity as the station-projection priors are varied. The Conservative case is dominated by broader limitations and is nearly flat. The Moderate and Optimistic cases become increasingly sensitive to weak global angular priors, while the projection sweep remains much less influential. The curves demonstrate explicitly that the high-sensitivity regime is prior limited.}
\label{fig:v13_prior_sweep}
\end{figure*}

The dependence on external angular information is shown in Figure~\ref{fig:v13_prior_sweep}. In the left panel the lunar-frame and lunar-ephemeris prior widths are co-scaled by the same factor; widening them eventually degrades the Moderate and especially the Optimistic forecast. The station-projection prior is swept separately in the right panel and is comparatively less influential over the displayed range.

\section{Sensitivity and systematic-control scale}
\label{sec:sensitivity}

The Fisher calculation above answers a conditional question: what uncertainty follows if the fitted model, covariance model, and stated priors are correct? A formal Fisher covariance alone does not test robustness to structured model misspecification. We therefore perform injection--recovery experiments in which the estimator remains fixed while additional residual structures are injected into the synthetic truth.

Accordingly, the GLS/joint-prior Fisher calculation provides the formal sensitivity forecast, while the Monte Carlo injection--recovery experiments assess its calibration and robustness across independent synthetic realizations under Matched and misspecified Stress conditions. Agreement in the Matched regime is interpreted as consistency with the formal Fisher behavior, not as proof of the model. Conversely, undercoverage and excess false positives in the Stress regime demonstrate where the formal covariance is insufficient for the declared mismatch budget.

\subsection{Matched and stress ensembles}

Two regimes are used. In the \emph{Matched} regime, the simulated truth contains the stated prior-controlled nuisance parameters, site-dependent colored noise, and data gaps, but no omitted deterministic mismatch. This regime tests calibration of the estimator itself. In the \emph{Stress} regime, the same fitted model is used but the truth additionally contains a declared synthetic mismatch budget.

The omitted-mismatch basis includes sine and cosine components for the lunar constituents M2, N2, and O1 and the solar constituents S2, K2, and P1; each pair spans first-order constituent amplitude--phase mismatch. Additional structures represent station- and constituent-specific ocean-loading mismatch, local long-period loading, distinct atmospheric and hydrological transfer residuals, deterministic network-common variability, stochastic cross-station covariance misspecification, calibration mismatch, quadratic nonstationarity, and polar-motion mismatch. Terrestrial-response uncertainty is carried inside the fitted model through one station-specific response gain and lag pair. The Stress truth deliberately exceeds that fitted basis by assigning independent gain, lag, and slow multiplicative drift perturbations to every station--constituent pair. For constituent reference signal \(g_{ic}^{\rm ref}(t)\), the added response mismatch is
\begin{equation}
\Delta g_{ic}^{\rm resp}(t)
=
\epsilon_{ic}g_{ic}^{\rm ref}(t)
+\tau_{ic}\frac{\dd g_{ic}^{\rm ref}}{\dd t}
+\kappa_{ic}\widetilde t\,g_{ic}^{\rm ref}(t),
\label{eq:response_stress_injection}
\end{equation}
where \(\widetilde t\) is centered and normalized within each station series. This construction tests unmodeled constituent dependence and temporal evolution of the effective terrestrial response rather than merely redrawing the fitted station-wide parameters. These process families are motivated by precision-gravimetry residual modeling \citep{VanCampVauterin2005,BoyHinderer2006,Neumeyer2010,Agnew2007}; their amplitudes are the declared synthetic budgets in Table~\ref{tab:v13_stress_budget}.

The constituent-level injections test a failure mode that a well-conditioned posterior Fisher matrix cannot reveal: over a finite observing interval, omitted amplitude--phase quadratures can project onto the slowly varying station geometry of the candidate signal. The Stress ensemble therefore asks whether the reported uncertainty remains calibrated under the declared mismatch families, rather than only whether \((\chi_1,\chi_2)\) are formally estimable.

No measured station residuals enter this calibration; the robust covariance is therefore conditional on the declared synthetic budgets. The code retains an optional interface for a future empirical calibration.

\begin{table*}[t]
\centering
\caption{\textbf{Declared synthetic omitted-mismatch budgets used in the Stress ensemble.}
All amplitude entries are in \(\mu{\rm Gal}\) except the dimensionless calibration fraction. Ocean loading is injected with station- and constituent-specific sine/cosine quadratures; the atmospheric and hydrological entries are separate transfer processes. ``Common'' is the deterministic network-common basis, whereas \(\sigma_{\rm net}\) is the RMS of the stochastic cross-station covariance. The polar value is the coefficient scale for each sine/cosine polar-mismatch component. These values define reproducible robustness scenarios and are not empirical station calibrations.}
\label{tab:v13_stress_budget}
\scriptsize
\setlength{\tabcolsep}{3.6pt}
\textbf{(a) Tidal, loading, environmental, and network processes}\\[3pt]
\begin{tabular}{lcccccccc}
\hline
Scenario & Lunar const. & Solar const. & Ocean load & Local load & Atmos. & Hydrol. & Common & \(\sigma_{\rm net}\)\\
\hline
Conservative & 0.0180 & 0.0135 & 0.0090 & 0.0090 & 0.00420 & 0.00420 & 0.00450 & 0.0300\\
Moderate     & 0.0066 & 0.0050 & 0.0033 & 0.0033 & 0.00155 & 0.00155 & 0.00165 & 0.0105\\
Optimistic   & 0.0034 & 0.00255& 0.0017 & 0.0017 & 0.00078 & 0.00078 & 0.00085 & 0.00375\\
\hline
\end{tabular}
\par\vspace{5pt}
\textbf{(b) Calibration, nonstationarity, polar, and terrestrial-response mismatch}\\[3pt]
\begin{tabular}{lcccccc}
\hline
Scenario & Calibration fraction & Nonstationary & Polar & Response gain & Response lag (s) & Response drift\\
\hline
Conservative & 0.0020 & 0.00450 & 0.00450 & 0.0025 & 30.0 & 0.00125\\
Moderate     & 0.0010 & 0.00165 & 0.00165 & 0.0010 & 10.0 & 0.00050\\
Optimistic   & 0.0005 & 0.00085 & 0.00085 & 0.0005 &  2.5 & 0.00025\\
\hline
\end{tabular}
\end{table*}

Each scenario/regime uses \(N_{\rm MC}=10^4\) Monte Carlo realizations. The ensemble is split into disjoint halves,
\begin{equation}
N_{\rm cal}=5000,
\qquad
N_{\rm val}=5000.
\label{eq:mc_split}
\end{equation}
The first half is used only to estimate the Monte-Carlo-calibrated covariance of the recovered two-component residual. The second half is reserved for out-of-sample validation of robust coverage and false-positive rates. The formal Fisher coverage and false-positive rates use the analytic posterior covariance and are evaluated over the full ensemble.

As a complementary isolation test, the gain, lag, and drift parts of Eq.~(\ref{eq:response_stress_injection}) are propagated separately through the fixed estimator. The combined response-only covariance inflates the worst-axis scale by factors \(1.13\), \(2.95\), and \(4.50\) in the Conservative, Moderate, and Optimistic scenarios, respectively; the lag perturbation is the largest single response contribution in the two more sensitive configurations. The full component-by-component values are reported in Table~\ref{tab:app_response_stress_impact}. This diagnostic is a conditional propagation of the declared synthetic response budget, not an empirical Love-number or observatory calibration.

For nonzero injections we use
\begin{equation}
(\chi_1,\chi_2)
=
(10^{-3},-5\times10^{-4}),
\label{eq:mc_injected_signal}
\end{equation}
while a separate null ensemble is used to determine false-positive behavior. Coverage is evaluated with the two-dimensional \(95\%\) ellipse threshold \(\Delta\chi^2=5.99146\).

\subsection{Injection--recovery results}

\begin{table*}[!htbp]
\centering
\caption{\textbf{Synthetic injection--recovery calibration and robustness tests.} The Matched regime contains the stated priors, site-dependent colored noise, and data gaps but no omitted deterministic mismatch. The Stress regime additionally injects the declared scenario-specific lunar--solar constituent, loading, atmospheric/hydrological-type environmental, network-common, terrestrial-response gain/lag/drift, calibration, nonstationary, and polar-mismatch budget. No stress amplitude is tuned to force nominal coverage. Formal coverage and false-positive rate (FPR) use the analytic joint-Fisher covariance and are evaluated on all Monte Carlo realizations. The Monte-Carlo-calibrated robust covariance is estimated only from the calibration half, while robust coverage and FPR are evaluated only on the independent validation half. The reported \(\sigma_{\rm robust}\) is the worst principal axis of the null-distribution covariance estimated from that calibration half. Panel (b) reports the mean recovery bias and root-mean-square error (RMSE) for both fitted residual coefficients.}
\label{tab:v13_mc_results}
\scriptsize
\setlength{\tabcolsep}{4.3pt}
\renewcommand{\arraystretch}{1.08}
\textbf{(a) Sensitivity, coverage, and false-positive diagnostics}\\[3pt]
\begin{tabular}{llccccccccc}
\hline
Scenario & Regime & \(N_{\rm cal}\) & \(N_{\rm val}\) & \(\sigma_{\rm formal}\) & \(\sigma_{\rm robust}\) & Infl. & Formal cov. & Robust cov. & Formal FPR & Robust FPR \\
\hline
Conservative & Matched & 5000 & 5000 & 9.513e-03 & 9.602e-03 & 1.01 & 0.950 & 0.953 & 0.052 & 0.052 \\
Conservative & Stress & 5000 & 5000 & 9.513e-03 & 1.142e-02 & 1.20 & 0.865 & 0.949 & 0.135 & 0.049 \\
Moderate & Matched & 5000 & 5000 & 4.490e-04 & 4.502e-04 & 1.00 & 0.951 & 0.952 & 0.053 & 0.051 \\
Moderate & Stress & 5000 & 5000 & 4.490e-04 & 1.360e-03 & 3.03 & 0.438 & 0.950 & 0.550 & 0.048 \\
Optimistic & Matched & 5000 & 5000 & 4.182e-05 & 4.239e-05 & 1.01 & 0.950 & 0.954 & 0.052 & 0.048 \\
Optimistic & Stress & 5000 & 5000 & 4.182e-05 & 1.972e-04 & 4.71 & 0.317 & 0.949 & 0.685 & 0.051 \\
\hline
\end{tabular}
\par\vspace{7pt}
\noindent\makebox[\textwidth][c]{%
  \textbf{(b) Bias and root-mean-square recovery error}}
\par\vspace{3pt}
\begin{tabular}{llcccc}
\hline
Scenario & Regime & Bias \(\chi_1\) & Bias \(\chi_2\) & RMSE \(\chi_1\) & RMSE \(\chi_2\) \\
\hline
Conservative & Matched & 1.511e-05 & -1.505e-07 & 1.558e-03 & 9.521e-03 \\
Conservative & Stress & -2.554e-05 & -7.437e-05 & 1.981e-03 & 1.131e-02 \\
Moderate & Matched & -1.245e-06 & -2.746e-06 & 1.723e-04 & 4.509e-04 \\
Moderate & Stress & -2.921e-06 & 6.279e-06 & 2.761e-04 & 1.366e-03 \\
Optimistic & Matched & -8.895e-08 & 1.038e-06 & 2.748e-05 & 4.170e-05 \\
Optimistic & Stress & -2.913e-07 & 4.552e-07 & 3.734e-05 & 2.003e-04 \\
\hline
\end{tabular}
\end{table*}

The Matched cases remain close to nominal \(95\%\) coverage and \(5\%\) ellipse false-positive rates. Under the response-expanded Stress construction, the formal covariance undercovers in every scenario, increasingly so as the nominal sensitivity improves: the formal coverages are \(0.865\), \(0.438\), and \(0.317\), with null ellipse false-positive rates \(0.135\), \(0.550\), and \(0.685\). Covariance estimated only from the independent calibration half restores held-out coverages to \(0.949\), \(0.950\), and \(0.949\), and held-out false-positive rates to \(0.049\), \(0.048\), and \(0.051\). For a validation fraction near \(0.95\) or \(0.05\) with \(N_{\rm val}=5000\), the binomial one-standard-error sampling scale is about \(0.0031\); the quoted rates should therefore be read as near-nominal Monte Carlo estimates rather than exact calibration identities. In the Optimistic--Stress case the robust worst-axis scale rises from the formal \(4.182\times10^{-5}\) to \(1.972\times10^{-4}\), an inflation factor of \(4.71\). Its recovery biases, \(-2.913\times10^{-7}\) and \(4.552\times10^{-7}\), remain small relative to RMSE values \(3.734\times10^{-5}\) and \(2.003\times10^{-4}\). Thus the dominant failure remains covariance underestimation and broadened recovery dispersion rather than a large systematic displacement of the recovered mean.

\begin{figure*}[t]
\centering
\includegraphics[width=0.96\textwidth]{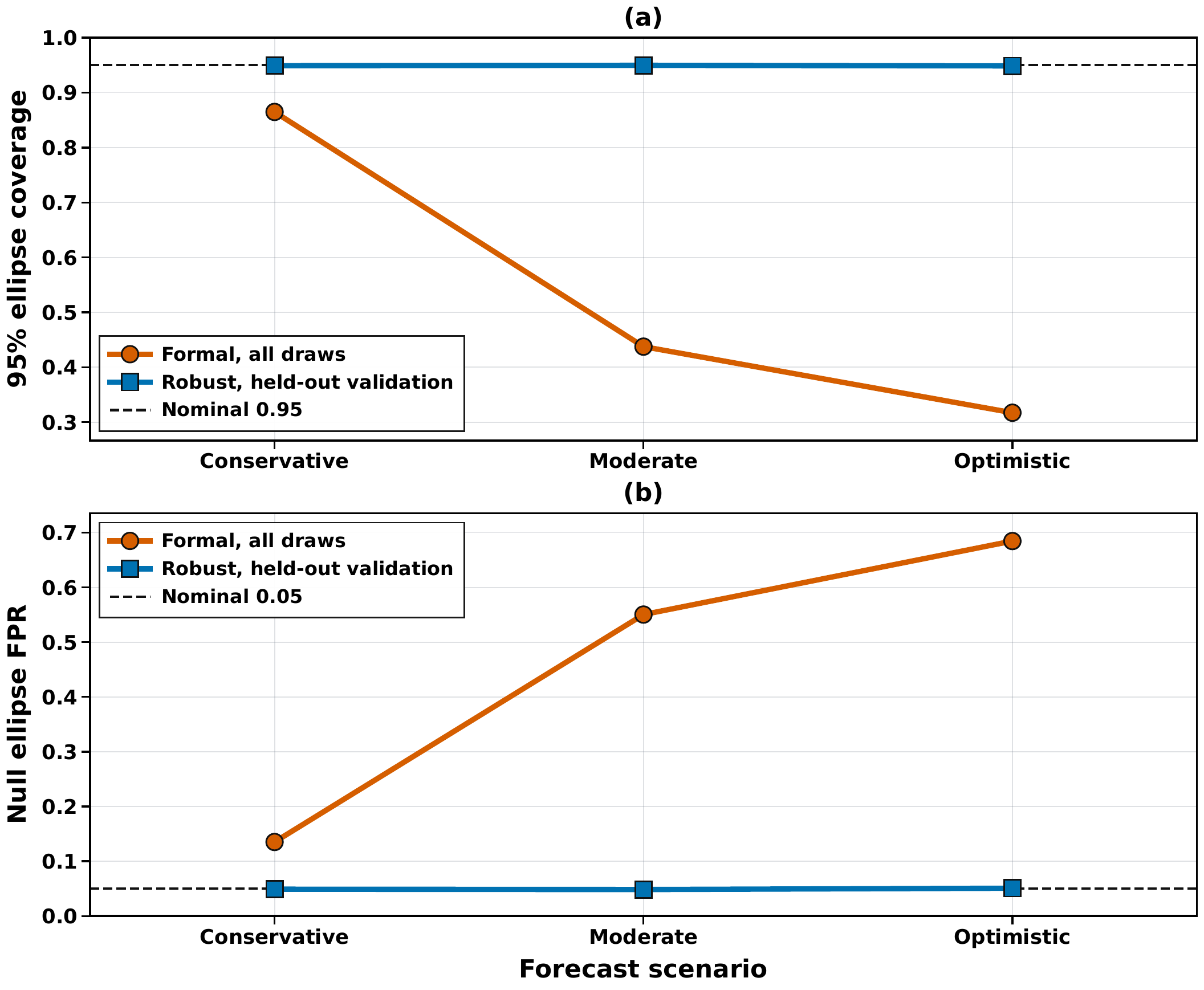}
\caption{\textbf{Out-of-sample validation under the response-expanded Stress construction.}
(a) The upper panel shows formal all-draw and robust held-out \(95\%\) joint-ellipse coverage. (b) The lower panel shows the corresponding null joint-ellipse false-positive rates. The formal covariance increasingly undercovers and over-rejects as the nominal forecast becomes more sensitive. The robust covariance is estimated only from the 5000-realization calibration half; its performance shown here is evaluated only on the disjoint 5000-realization validation half. The near-nominal blue curves therefore represent out-of-sample calibration, not reuse of the training ensemble.}
\label{fig:v13_stress_validation}
\end{figure*}

\begin{figure*}[t]
\centering
\includegraphics[width=0.90\textwidth]{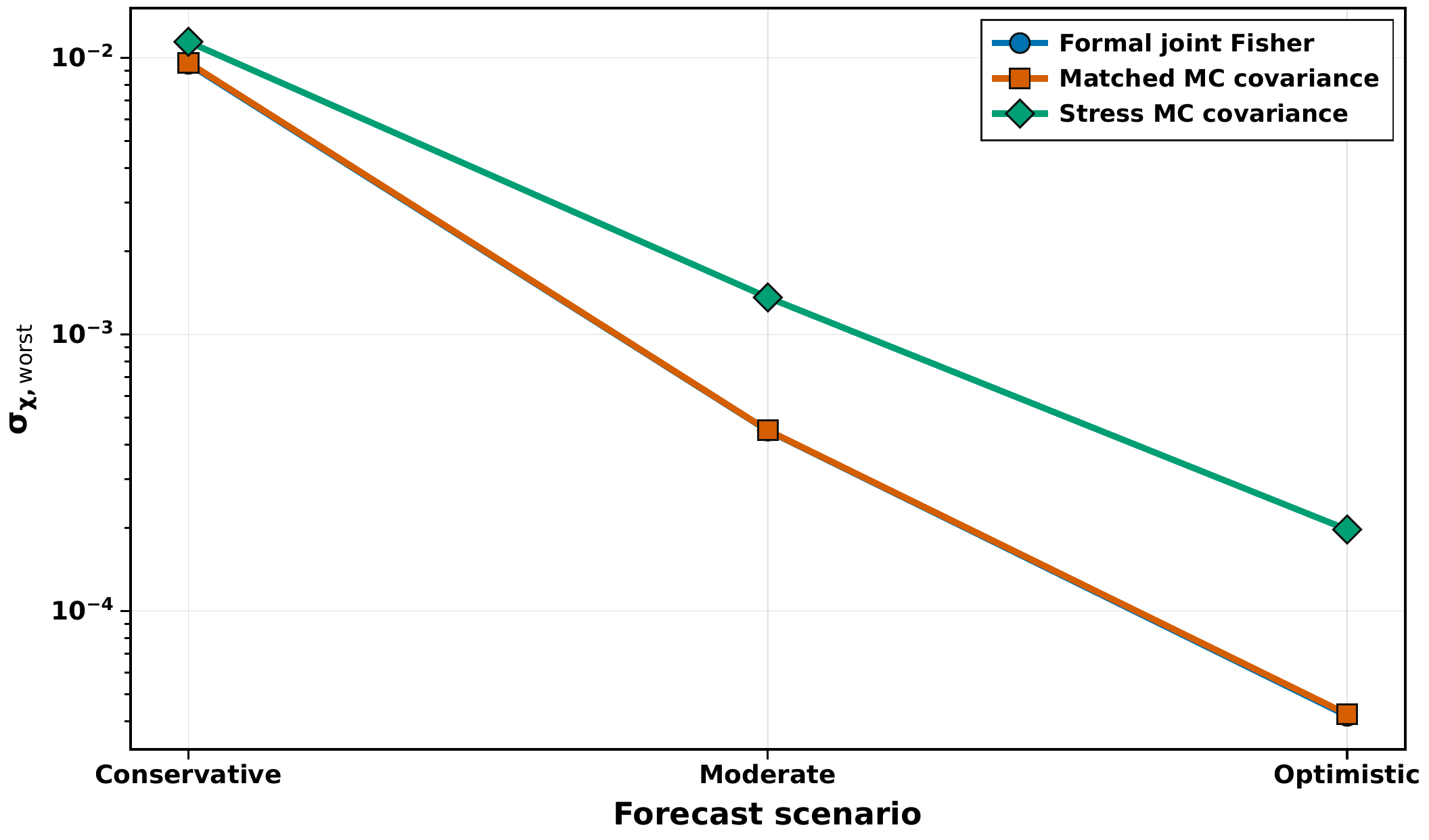}
\caption{\textbf{Formal joint-Fisher sensitivity versus Monte-Carlo-calibrated robustness floor.}
 The formal curve uses the analytic joint-Fisher covariance, while the Matched and Stress MC curves use \(\sigma_{\rm robust}\), defined as the worst principal axis of the null-distribution covariance estimated from the 5000-realization calibration half. Matched scales remain close to the formal values in all scenarios. The response-expanded Stress scales show increasing inflation, reaching \(1.972\times10^{-4}\) relative to the formal \(4.182\times10^{-5}\) in the Optimistic scenario. These are synthetic Monte-Carlo covariance scales, not empirical superconducting-gravimeter measurements; robust coverage and FPR are evaluated separately on the independent validation half.}
\label{fig:v13_robust_floor}
\end{figure*}

\subsection{Relation to the Halilsoy morphology}

The numerical forecast constrains only the generic coefficients \(\chi_1,\chi_2\). It does not fit \(\gamma_H\). A model-dependent \(\gamma_H\) forecast would require a fully specified, finite global signal template with the complete time dependence \(a_0(t)\eta_H(\alpha,t,\rho)\) and the same three-dimensional station projection. Near zeros of the Halilsoy plus-sector factor, the ratio \(\eta_H=3C_H/P_H\) is not a globally well-conditioned amplitude variable. For this reason the present forecast does not impose \(\chi_A=\gamma_H\eta_H\) as a global estimator identity.


\section{Discussion}
\label{sec:discussion}

The purpose of this formulation is to make the physical status of the proposed sine-quadrature channel explicit. The result is not a new leading-order theory of lunar tides. The dominant Earth--Moon tide is already described by the Newtonian quadrupolar tensor and by standard geophysical tide modeling. The present work instead defines a residual tensor channel that could be estimated or bounded after the ordinary lunar, solar, terrestrial, environmental, instrumental, and reference-frame contributions have been modeled.

The sensitivity estimates in Sections~\ref{sec:observation_model} and~\ref{sec:sensitivity} should be read as illustrative forecasts, not as a real-data detection. The calculation now specifies one globally coherent three-dimensional tensor, a joint-prior treatment of the exact signal--frame degeneracy, structured lunar--solar and geophysical mismatch, correlated network noise, and independent calibration and validation ensembles. It therefore provides a quantitative separability and robustness test under stated synthetic assumptions. A full observational analysis would still require real residual data, precision ephemerides, station-specific loading and response products, and an empirically justified covariance model.

\subsection{From a formal ansatz to a residual-estimation framework}

The local tensor introduced earlier should be interpreted as a phenomenological residual operator, not as a derived Earth--Moon metric solution. It identifies the angular channel in which a symmetric off-diagonal residual would appear and provides an estimable or boundable parameterization. The operational forecast extends that local basis to the coherent three-dimensional system in Eqs.~(\ref{eq:cross_tensor_3d})--(\ref{eq:joint_linear_model_3d}), projects the same two-component tensor onto every station vertical, and carries the resulting degeneracies into the joint-prior Fisher problem.

This distinction is important. A theoretical prediction of a nonzero residual would require a self-consistent source or matching calculation that derives an off-diagonal Earth--Moon tidal component from a physical metric, post-Newtonian correction, modified source structure, or environmental coupling. The present paper does not claim such a derivation. Its contribution is to supply the residual basis, the exact identifiability result, and the estimator and validation structure needed before such a coefficient could be interpreted.

\subsection{Why the sine channel is necessary but not sufficient}

The sine-quadrature channel is necessary for a symmetric off-diagonal residual in the local model, but it is not sufficient for a unique physical interpretation. The first-order three-dimensional calculation strengthens the original frame-misregistration argument: each lunar-frame derivative column is exactly three times the corresponding cross-residual column. The data-only Fisher matrix is therefore singular in those directions, and a finite constraint requires independent frame or ephemeris information.

This degeneracy is not a small presentational caveat. It is the central systematic limitation of the proposed channel. Frame, projection, ephemeris, loading, terrestrial-response, environmental, instrumental, site-dependent, and network-correlated terms must therefore enter the fit or its covariance model. The prior-width sweep and the held-out Stress tests make the two consequences visible: external angular information controls the formal identifiability, while omitted structured mismatch controls calibration and false-positive behavior as nominal precision improves.

\subsection{Relation to the original \texorpdfstring{\(45^\circ\)}{45-degree} tidal idea}

The original motivation for this work was the intuition that a cross-polarized gravitational structure should produce a \(45^\circ\)-type tidal signature. 
The present formulation keeps that intuition but translates it into tensor language. 
The ordinary diagonal Newtonian tide has a perfectly valid projection at \(45^\circ\):
\begin{equation}
a_N(45^\circ)
=
\frac12a_0.
\label{eq:discussion_newtonian_45}
\end{equation}
This value is not an anomaly and not a cross residual.

The independent residual target is instead the sine-quadrature component
\begin{equation}
a_{\rm cross}(\beta)
=
a_0\chioff\sin(2\beta),
\label{eq:discussion_cross_45}
\end{equation}
whose extrema occur along the \(45^\circ\)-type directions. 
Thus the phrase \(45^\circ\)-type residual refers to the angular extrema of the off-diagonal basis and to the rotation of the local eigenframe, not to the ordinary Newtonian projection along a \(45^\circ\) line.

In the global forecast the same morphology is represented by two transverse coefficients and the full station projection; it is not implemented by assigning each station an independent two-dimensional angle.

\subsection{Halilsoy-inspired morphology and normalization}

The Halilsoy sector is used in this paper only as a local relativistic morphology template. 
It supplies an explicit example of an off-diagonal tidal block in which a cross-polarized structure rotates a local eigenframe. 
It does not supply a normalization for the Earth--Moon residual amplitude. 
The forecast therefore constrains the generic coefficients \(\chi_1\) and \(\chi_2\), not \(\gamma_H\). A direct \(\gamma_H\) analysis would require a finite global template with the complete time dependence \(a_0(t)\eta_H(\alpha,t,\rho)\) and the same three-dimensional station projection.

This separation also prevents the cross-to-plus ratio near zeros of the Halilsoy plus sector from being interpreted as a divergent physical lunar acceleration. The finite geometric variable is the eigenframe angle; the finite observational variables are the projected cross coefficients and their acceleration response.

\subsection{Perturbative eigenframe interpretation}

The eigenframe language remains valid in the perturbative residual regime. The eigenvalue shifts of the local symmetric tensor are second order in the off-diagonal coefficient, whereas the principal-frame rotation is first order. The model is therefore not merely a rigid rotation of the Newtonian tensor at finite amplitude. It is a perturbative residual tensor whose leading local effect is naturally represented by the sine-quadrature coefficient and whose global measurement requires the full three-dimensional projection.

\subsection{Limits of the two-dimensional representation}

The two-dimensional formulae define the angular basis in a local plane; they are not an operational Earth-tide model. The three-dimensional forecast includes the local vertical projection, ephemeris and frame controls, terrestrial response, constituent-specific ocean loading, atmosphere, hydrology, polar motion, station terms, and cross-station stochastic covariance. No numerical conclusion therefore depends on assuming that the Moon lies in one station-specific vertical--north plane.

\subsection{What would count as evidence}

A discovery-level result would not be a generic mismatch in a tidal record. It would require a statistically significant nonzero two-component residual that remains after removal or marginalization of the Newtonian lunar and solar tides, solid-Earth response, ocean and local loading, atmospheric and hydrological effects, polar motion, local site effects, instrumental drift, calibration error, and frame/projection uncertainties. It would also require calibrated coverage and false-positive behavior under mismatch families justified independently for the selected network.

For a future application to superconducting-gravimeter or related precision-gravity data, the conventional tide-generating-potential, Earth-orientation, solid-Earth, and station-loading models should be constructed first \citep{Doodson1921,CartwrightEdden1973,Tamura1987,HartmannWenzel1995,PetitLuzum2010,Agnew2007}. The two cross-residual columns would then be appended to the residual design matrix. Synthetic priors and Stress amplitudes used here must be replaced by documented station products and measured post-correction covariance, not merely retuned until a preferred sensitivity is obtained.

\subsection{Back-action and dynamical closure}

The present paper treats the off-diagonal sector as a residual coefficient in a local tidal operator. 
It does not solve the dynamical back-action problem. 
A complete physical theory would need to determine whether such a residual sector can be coupled self-consistently to the background curvature, Earth--Moon orbital dynamics, and terrestrial response. 
Schematically, such a program would require
\begin{equation}
\E_{\rm cross}
\longrightarrow
\delta g_{\mu\nu}
\longrightarrow
\delta R_{\mu\nu\rho\sigma}
\longrightarrow
\delta \mathcal{O}_{\rm EM}
\longrightarrow
\delta a_{\rm obs}.
\label{eq:discussion_back_action_chain}
\end{equation}
Here \(\delta g_{\mu\nu}\) is a metric perturbation, \(\delta R_{\mu\nu\rho\sigma}\) the corresponding curvature perturbation, \(\delta \mathcal{O}_{\rm EM}\) the perturbation of the Earth--Moon orbital and geophysical response system, and \(\delta a_{\rm obs}\) the observable acceleration residual.

This paper does not claim to derive \(\gamH\) from such a closed dynamical model. 
The coefficient \(\gamH\) is therefore phenomenological. 
Future work would need either a self-consistent gravitational--orbital--geophysical derivation, or a real-data residual analysis that constrains the corresponding finite time-dependent Halilsoy-inspired template through the estimator described above.

The absence of such a derivation or data analysis is therefore a limitation of the present work, but it does not make the residual channel mathematically empty. 
The framework developed here specifies what would have to be fitted, what degeneracies must be controlled, and what level of systematic error would have to be reached before an off-diagonal lunar residual could be interpreted physically. 
In this sense, the present paper provides a theoretical and statistical target for future tests, rather than a proof that such a residual is already present in the Earth--Moon system.

\subsection{Summary of the interpretation}

The interpretation adopted here is conservative. 
First, no observational detection is claimed. 
Second, the Halilsoy solution is used only as a local morphology template, not as a global Earth--Moon model. 
Third, the two-dimensional tensor formulae define an angular residual basis, not a complete gravimetric forward model. 
Fourth, the sine-quadrature coefficient is degenerate with frame, projection, loading, environmental, instrumental, and site-response effects. 
Fifth, the dynamical back-action and amplitude-normalization problem remains open.

\textbf{The contribution is a methodological residual-estimation and forecasting framework that makes signal identifiability, prior dependence, and robustness to model mismatch explicit. The reported sensitivities are conditional on the declared synthetic assumptions; they establish neither a detection nor a new Earth--Moon gravitational mechanism and do not demonstrate the observing capability of a real gravimeter network.}

The coherent three-dimensional tensor, exact prior-limited degeneracy, response-expanded Stress generator, network covariance, and held-out calibration test extend that original residual basis into a reproducible inference problem.

Three limitations delimit the numerical claims. The loading, response, environmental, and network-covariance budgets are synthetic rather than empirical station characterizations. The Stress ensemble is a finite collection of mismatch families, and both halves of the calibration--validation split are generated from that declared family. Finally, network size, duration, cadence, noise, and prior quality co-vary across scenarios, so their ordering cannot isolate a Halilsoy-specific, station-number, or instrumental gain. These restrictions do not alter the exact three-dimensional identifiability result, but they make every finite sensitivity and robust floor conditional on the stated priors and synthetic budgets.


\section{Conclusion}
\label{sec:conclusion}

We have developed a three-dimensional, frame-degeneracy-aware sine-quadrature framework for estimating or bounding possible off-diagonal lunar tidal residuals. The starting point remains the standard Newtonian quadrupolar tidal tensor, which describes the dominant Earth--Moon tide in its local principal frame. A projection of that diagonal Newtonian field at \(45^\circ\) is possible, but it is not an independent cross-tidal residual. The residual channel considered here is instead the off-diagonal sine-quadrature sector, represented globally by two transverse coefficients and projected from one coherent tensor onto every station vertical.

The central identifiability result is
\begin{equation}
\left.
\frac{\partial g_N^M}{\partial\delta_A}
\right|_{\delta=0}
=
3
\frac{\partial \Delta g^{\rm cross}}{\partial\chi_A},
\end{equation}
which makes the leading lunar-frame degeneracy exact at first order. Accordingly, the data-only signal/control Fisher block is singular and a finite bound on the cross-residual coefficients is intrinsically prior limited.

This result preserves the original \(45^\circ\)-type motivation while stating it more precisely. The \(45^\circ\)-type directions are the extrema of the local off-diagonal basis and characterize the associated eigenframe rotation; they are not directions along which ordinary Newtonian gravity is absent. A nonzero sine coefficient is therefore not by itself evidence for a new lunar tidal sector. It must be interpreted only after frame, projection, ephemeris, terrestrial-response, loading, environmental, instrumental, site-dependent, and correlated-network residuals have been modeled, bounded, or marginalized.

\textbf{For the three synthetic scenarios, the joint-prior worst-axis uncertainties in the dimensionless coefficients \(\chi_1\) and \(\chi_2\) are}
\begin{equation}
9.513\times10^{-3},
\qquad
4.490\times10^{-4},
\qquad
4.182\times10^{-5},
\end{equation}
for the Conservative, Moderate, and Optimistic configurations, respectively. The canonical signal--nuisance diagnostic gives \(R^2_{\rm all}=1\) in every case, while \(R^2_{\rm geo}\) decreases from \(0.195\) to \(0.091\) to \(0.073\).

Matched injection--recovery ensembles are calibrated at approximately the nominal \(95\%\) coverage and \(5\%\) ellipse false-positive levels. Under the response-expanded declared Stress budget, the formal covariance undercovers in all three scenarios, with the strongest effect in the Optimistic case. There the formal uncertainty of \(4.182\times10^{-5}\) gives \(0.317\) coverage and a \(0.685\) ellipse false-positive rate, whereas a null-distribution covariance calibrated on the independent calibration half yields a worst-axis scale of \(1.972\times10^{-4}\), with held-out coverage \(0.949\) and ellipse false-positive rate \(0.051\). The corresponding mean biases remain small relative to the recovery RMSE, indicating that the dominant Stress effect is covariance broadening rather than a large mean shift. The separate response-only propagation identifies constituent-dependent phase lag as the largest individual terrestrial-response contribution under the declared Moderate and Optimistic budgets.

The Halilsoy cross-polarized sector is retained only as a local morphology template. The primary forecast parameters are \(\chi_1,\chi_2\); no direct bound on \(\gamma_H\) is claimed. A \(\gamma_H\) analysis would require a separate finite, time-dependent matched signal model and a physical normalization linking the local morphology to the Earth--Moon system.

\textbf{These results provide a quantitative basis for designing future gravimetric residual searches: they identify the external constraints and robustness checks required for a physically interpretable bound.} The result is not a detection or a replacement for standard lunar tidal theory. Its contribution is to define a specific off-diagonal residual channel, derive its local sine-quadrature geometry, extend it to a coherent three-dimensional station response, expose its exact frame degeneracy, separate Halilsoy morphology from physical amplitude, and test the resulting estimator against declared structured mismatch. The quoted scenario improvements and Monte-Carlo-calibrated floors are model-specific rather than measured or universal limits; an observational application requires station-specific calibration and independent external frame information.

Future work should proceed along two complementary directions. The estimator should first be applied to precision gravimetric or geodetic residuals using standard terrestrial tide, loading, Earth-orientation, instrumental, and site-response products with an empirically measured network covariance. In parallel, the dynamical back-action and amplitude-normalization problem must be developed to determine whether a parameter such as \(\gamma_H\) can be derived from a self-consistent gravitational--orbital--geophysical perturbation theory. Until those steps are completed, the present result is best understood as a reproducible theoretical and statistical target for a controlled residual search.


\begin{acknowledgments}

The author is deeply grateful to Professor Mustafa Halilsoy for fruitful discussions, for drawing attention to possible \(45^\circ\)-type tidal signatures associated with gravitational-wave cross polarization, and for formulating the original physical question that motivated this work. 
The author also thanks him for his guidance, encouragement, and insightful suggestions during the development of the present residual framework. 
This work grew out of the attempt to translate that physical intuition into a tensorial, harmonic, and observationally testable mathematical form.

The author also thanks the anonymous referee for a careful reading and constructive comments which helped improve the clarity, scope, and quantitative content of the revised manuscript.

\end{acknowledgments}


\section*{Data Availability}

\textbf{This study uses only synthetic data generated by the Julia implementation; no observational gravimeter data are analyzed. Numerical inputs and principal results are presented in Appendix~\ref{app:numerical_data}, and the complete source code is reproduced in Appendix~\ref{app:computational_source}. The standalone code and reference outputs are archived on Zenodo \citep{Eser2026ForecastSoftware}, \doi{10.5281/zenodo.22896632}, and the code is also available on \href{https://github.com/muhittineser-lang/halilsoy-lunar-residual-forecast}{GitHub}. The default calculation requires no personal paths or external data. Reproducibility verifies the calculations under the declared synthetic assumptions, not their empirical validity.}


\clearpage
\bibliography{moon_cross_tidal_references_v3}
\bibliographystyle{aasjournalv7}
\clearpage

\appendix
\suppressfloats[t]

\section{Numerical data underlying the synthetic forecast}
\label{app:numerical_data}

This appendix records the Julia inputs and principal outputs underlying Sections~\ref{sec:observation_model}--\ref{sec:sensitivity}, making the synthetic assumptions and reported results directly reproducible.

\subsection{Physical constants, time conventions, and signal normalization}

The Julia calculation uses the constants and periods listed in Table~\ref{tab:app_constants}. Distances are in meters, masses in kilograms, and time periods in days unless otherwise stated. The conversion \(1~{\rm m\,s^{-2}}=10^8~\mu{\rm Gal}\) is used. The quantity \(55~\mu{\rm Gal}\) is retained only as the benchmark scale used in the local explanatory figures; the numerical forecast uses the time-dependent \(a_0(t)=GM_MR_E/D_M^3(t)\).

\begin{table*}[t]
\centering
\caption{\textbf{Physical constants and time conventions used by the Julia synthetic calculation.}
The lunar and solar distance laws are simple deterministic surrogates used for the methodological forecast, not precision ephemerides.}
\label{tab:app_constants}
\footnotesize
\setlength{\tabcolsep}{5.0pt}
\begin{tabular}{lll}
\hline
Quantity & Symbol / implementation value & Numerical value\\
\hline
Newtonian constant & \(G\) & \(6.67430\times10^{-11}\ {\rm m^3\,kg^{-1}\,s^{-2}}\)\\
Lunar mass & \(M_M\) & \(7.342\times10^{22}\ {\rm kg}\)\\
Solar mass & \(M_S\) & \(1.98847\times10^{30}\ {\rm kg}\)\\
Earth radius & \(R_E\) & \(6.371\times10^{6}\ {\rm m}\)\\
Mean lunar orbital scale & \(A_M\) & \(3.844\times10^{8}\ {\rm m}\)\\
Lunar eccentricity surrogate & \(e_M\) & \(0.0549\)\\
Astronomical unit & AU & \(1.495978707\times10^{11}\ {\rm m}\)\\
Earth-orbit eccentricity surrogate & \(e_\oplus\) & \(0.0167086\)\\
Sidereal day & \(T_{\rm sid}\) & \(0.99726968\ {\rm d}\)\\
Lunar sidereal period & \(T_{M,\rm sid}\) & \(27.321661\ {\rm d}\)\\
Lunar anomalistic period & \(T_{M,\rm anom}\) & \(27.55455\ {\rm d}\)\\
Year & \(T_{\rm yr}\) & \(365.256363004\ {\rm d}\)\\
Fortnightly nuisance period & \(T_{\rm fort}\) & \(14.77\ {\rm d}\)\\
Monthly nuisance period & \(T_{\rm month}\) & \(27.321661\ {\rm d}\)\\
Polar-motion period & \(T_{\rm polar}\) & \(433.0\ {\rm d}\)\\
Acceleration conversion & \(1~{\rm m\,s^{-2}}\) & \(10^8~\mu{\rm Gal}\)\\
Local benchmark only & \(a_{0,\rm bench}\) & \(55.0~\mu{\rm Gal}\)\\
\hline
\end{tabular}
\end{table*}

The distance surrogates used in the numerical calculation are
\begin{align}
D_M(t)
&=
A_M\left[1-e_M\cos\!\left(\frac{2\pi t}{T_{M,\rm anom}}\right)\right],
\\
D_S(t)
&=
{\rm AU}\left[1-e_\oplus\cos\!\left(\frac{2\pi t}{T_{\rm yr}}\right)\right].
\end{align}
These expressions define the time-dependent lunar and solar scales used only in the synthetic forecast. In a real-data analysis they would be replaced by precision ephemerides and conventional Earth-orientation transformations.

\subsection{Scenario inputs, priors, gaps, and random seeds}

Table~\ref{tab:app_scenario_controls} gives the numerical controls that supplement Table~\ref{tab:v13_assumptions}. The random-gap fraction is \(0.03\) in every scenario, and one finite outage of 12 hours is generated independently for each station. The station coordinates are those given in Eq.~\ref{eq:v13_station_sets}.

\begin{table*}[t]
\centering
\caption{\textbf{Julia scenario controls not fully collected in the main scenario table.}
Response gain is dimensionless; loading and polar widths are in \(\mu{\rm Gal}\).}
\label{tab:app_scenario_controls}
\scriptsize
\setlength{\tabcolsep}{3.5pt}
\begin{tabular}{lcccccccc}
\hline
Scenario & Response gain & Response lag (s) & Local load prior & Polar prior &
Gap fraction & Outages/station & Outage (h) & Gap seed\\
\hline
Conservative & 0.005 & 60 & 0.0090 & 0.0050 & 0.03 & 1 & 12 & 101\\
Moderate     & 0.002 & 20 & 0.0033 & 0.0020 & 0.03 & 1 & 12 & 202\\
Optimistic   & 0.001 &  5 & 0.0017 & 0.0010 & 0.03 & 1 & 12 & 303\\
\hline
\end{tabular}
\end{table*}

The fitted and Stress process scales are collected in Table~\ref{tab:app_process_controls}. Ocean entries are total station-level RMS scales distributed across the six constituent sine/cosine pairs; atmosphere and hydrology remain separate station-specific transfer processes.

\begin{table*}[t]
\centering
\caption{\textbf{Fitted-prior and Stress process scales in \(\mu{\rm Gal}\).}
Each entry is shown as fitted/Stress. Network denotes the RMS scale of the stochastic cross-station covariance.}
\label{tab:app_process_controls}
\scriptsize
\setlength{\tabcolsep}{4.5pt}
\begin{tabular}{lccccc}
\hline
Scenario & Ocean loading & Local loading & Atmosphere & Hydrology & Network\\
\hline
Conservative & 0.0090/0.0090 & 0.0090/0.0090 & 0.0060/0.0042 & 0.0060/0.0042 & 0.0200/0.0300\\
Moderate     & 0.0033/0.0033 & 0.0033/0.0033 & 0.0022/0.00155 & 0.0022/0.00155 & 0.0070/0.0105\\
Optimistic   & 0.0017/0.0017 & 0.0017/0.0017 & 0.0011/0.00078 & 0.0011/0.00078 & 0.0025/0.00375\\
\hline
\end{tabular}
\end{table*}

The site-dependent colored-noise construction starts from the scenario mean \((\sigma_n,\rho)\) values in Table~\ref{tab:v13_assumptions}. Across the stations of a multi-station scenario, the Julia calculation applies a deterministic linear spread of \(\pm15\%\) to the innovation scale and a deterministic spread of \(\pm0.05\) to the AR(1) coefficient, with the resulting \(\rho\) values clipped to the interval \([0,0.95]\). For the single-station Conservative case, the listed scenario values are used directly.

The Monte Carlo calculation uses a base random seed of \(12345\), with the scenario gap seed added internally. The Stress regime uses an additional seed offset of \(100000\) so that its random stream is distinct from the Matched regime. Each scenario/regime contains \(10^4\) realizations, split deterministically into the first 5000 calibration realizations and the final 5000 validation realizations. The nonzero injected residual is
\begin{equation}
(\chi_1,\chi_2)=(10^{-3},-5\times10^{-4}),
\end{equation}
and a separate null ensemble uses \((\chi_1,\chi_2)=(0,0)\).

\subsection{Constituent and stress-model numerical inputs}

The constituent periods used to generate the omitted tidal-mismatch basis are listed in Table~\ref{tab:app_constituents}. Each named constituent is represented by sine and cosine columns, so that small amplitude and phase errors are both spanned at first order.

\begin{table}[t]
\centering
\caption{\textbf{Constituent periods used in the Julia Stress basis.}}
\label{tab:app_constituents}
\footnotesize
\begin{tabular}{lcc}
\hline
Constituent & Family & Period (h)\\
\hline
M2 & lunar & 12.4206012\\
N2 & lunar & 12.6583475\\
O1 & lunar & 25.8193387\\
S2 & solar & 12.0000000\\
K2 & solar & 11.9672348\\
P1 & solar & 24.0658877\\
\hline
\end{tabular}
\end{table}

The declared scenario-level Stress amplitudes are repeated in Table~\ref{tab:app_stress_full} so that the numerical data appendix is self-contained. They are synthetic controls, not measured residual amplitudes.

\begin{table*}[t]
\centering
\caption{\textbf{Declared synthetic Stress budgets used in the Julia injection--recovery calculation.}
All entries are in \(\mu{\rm Gal}\). Atmospheric and hydrological entries are separate process amplitudes; \(\sigma_{\rm net}\) is the stochastic network RMS.}
\label{tab:app_stress_full}
\scriptsize
\setlength{\tabcolsep}{3.6pt}
\begin{tabular}{lcccccccc}
\hline
Scenario & Lunar const. & Solar const. & Ocean load & Local load & Atmos. & Hydrol. & Common & \(\sigma_{\rm net}\)\\
\hline
Conservative & 0.0180 & 0.0135 & 0.0090 & 0.0090 & 0.00420 & 0.00420 & 0.00450 & 0.0300\\
Moderate     & 0.0066 & 0.0050 & 0.0033 & 0.0033 & 0.00155 & 0.00155 & 0.00165 & 0.0105\\
Optimistic   & 0.0034 & 0.00255& 0.0017 & 0.0017 & 0.00078 & 0.00078 & 0.00085 & 0.00375\\
\hline
\end{tabular}
\end{table*}

The fitted atmospheric periods are \(2.7\), \(5.3\), and \(9.1\) days and the Stress periods are \(2.9\), \(5.8\), and \(8.4\) days. The corresponding hydrological periods are \(29.5\), \(61\), and \(120\) days in the fitted model and \(34\), \(73\), and \(145\) days in the Stress generator. The deterministic network-common basis uses sine/cosine pairs with periods \(2.5\), \(7\), \(20\), and \(60\) days. The stochastic Stress network instead uses \(2.3\), \(7.49\), \(17.8\), and \(67.2\) days; its correlation strengths are \(0.45\), \(0.50\), and \(0.55\), with angular lengths \(52.5^\circ\), \(45^\circ\), and \(37.5^\circ\) for the three scenarios. Calibration fractions are \(0.002\), \(0.001\), and \(0.0005\); nonstationary and polar-mismatch amplitudes are \(0.00450\), \(0.00165\), and \(0.00085\,\mu{\rm Gal}\). The nonstationary component is a station-wise normalized quadratic drift. The polar-mismatch basis uses a \(430\)-day period in the Stress generator, while the fitted prior-controlled polar term uses the \(433\)-day reference period listed in Table~\ref{tab:app_constants}.

The constituent-specific terrestrial-response Stress generator adds three independent columns for every station and each of the six constituents in Table~\ref{tab:app_constituents}: a fractional gain perturbation, a phase-lag perturbation in seconds, and a fractional slow-drift perturbation. This gives 18, 54, and 180 response-mismatch columns in the one-, three-, and ten-station scenarios. The respective \((\sigma_{\rm gain},\sigma_{\rm lag},\sigma_{\rm drift})\) budgets are \((0.0025,30\,{\rm s},0.00125)\), \((0.0010,10\,{\rm s},0.00050)\), and \((0.0005,2.5\,{\rm s},0.00025)\).

\begin{table*}[t]
\centering
\caption{\textbf{Isolated terrestrial-response Stress propagation.}
For each component, \(\sigma_{\rm resp}\) is the worst principal axis generated by that response mismatch alone; \(\sigma_{\rm comb}\) combines it with the formal covariance, and ``Infl.'' is \(\sigma_{\rm comb}/\sigma_{\rm formal}\). These are conditional synthetic propagation diagnostics, not measured terrestrial-response uncertainties.}
\label{tab:app_response_stress_impact}
\footnotesize
\begin{tabular}{llccc}
\hline
Scenario & Response component & \(\sigma_{\rm resp}\) & \(\sigma_{\rm comb}\) & Infl.\\
\hline
Conservative & Gain  & 2.221e-03 & 9.764e-03 & 1.026\\
             & Lag   & 3.893e-03 & 1.028e-02 & 1.080\\
             & Drift & 2.029e-03 & 9.727e-03 & 1.022\\
             & All   & 4.910e-03 & 1.071e-02 & 1.125\\
Moderate     & Gain  & 2.587e-04 & 5.151e-04 & 1.147\\
             & Lag   & 1.196e-03 & 1.277e-03 & 2.845\\
             & Drift & 2.512e-04 & 5.145e-04 & 1.146\\
             & All   & 1.247e-03 & 1.326e-03 & 2.952\\
Optimistic   & Gain  & 6.444e-05 & 7.676e-05 & 1.835\\
             & Lag   & 1.662e-04 & 1.713e-04 & 4.097\\
             & Drift & 4.492e-05 & 6.137e-05 & 1.467\\
             & All   & 1.837e-04 & 1.884e-04 & 4.505\\
\hline
\end{tabular}
\end{table*}

\subsection{Joint-prior Fisher outputs}

The principal Julia-generated joint-prior outputs are listed in Table~\ref{tab:app_fisher_outputs}. These are the numerical values plotted in the full joint-prior comparison and the nuisance-subspace diagnostic, and they are identical to the values reported in the main text.

\begin{table*}[t]
\centering
\caption{\textbf{Principal joint-prior Fisher outputs generated by the Julia calculation.}}
\label{tab:app_fisher_outputs}
\footnotesize
\begin{tabular}{lcccccc}
\hline
Scenario & \(\sigma(\chi_1)\) & \(\sigma(\chi_2)\) & \(\sigma_{\rm worst}\) &
\(R^2_{\rm all}\) & \(R^2_{\rm geo}\) & Used fraction\\
\hline
Conservative & \(1.567\times10^{-3}\) & \(9.512\times10^{-3}\) & \(9.513\times10^{-3}\) & 1.000 & 0.195 & 0.956\\
Moderate     & \(1.742\times10^{-4}\) & \(4.490\times10^{-4}\) & \(4.490\times10^{-4}\) & 1.000 & 0.091 & 0.963\\
Optimistic   & \(2.736\times10^{-5}\) & \(4.182\times10^{-5}\) & \(4.182\times10^{-5}\) & 1.000 & 0.073 & 0.968\\
\hline
\end{tabular}
\end{table*}

For the prior-width study, the Julia calculation uses 25 logarithmically spaced scale factors
\begin{equation}
s_k=10^{-1+2(k-1)/24},
\qquad
k=1,\ldots,25.
\label{eq:app_prior_grid}
\end{equation}
In the left panel of Figure~\ref{fig:v13_prior_sweep}, the baseline lunar-frame and lunar-ephemeris prior widths are multiplied by the same \(s_k\) while the remaining groups are held fixed. In the right panel only the station-projection prior widths are multiplied by \(s_k\). The plotted curves are direct Julia outputs on this grid.

\subsection{Injection--recovery numerical outputs}

Table~\ref{tab:app_mc_outputs_a} records the formal and held-out robust diagnostics, and Table~\ref{tab:app_mc_outputs_b} records the corresponding nonzero-injection bias and RMSE values. These are the numerical data underlying the main injection--recovery and formal-versus-robust comparisons.

\begin{table*}[t]
\centering
\caption{\textbf{Julia injection--recovery sensitivity, coverage, and false-positive outputs.}
The robust covariance is estimated from the 5000-realization calibration half and validated on the independent 5000-realization half.}
\label{tab:app_mc_outputs_a}
\scriptsize
\setlength{\tabcolsep}{3.2pt}
\begin{tabular}{llccccccccc}
\hline
Scenario & Regime & \(N_{\rm cal}\) & \(N_{\rm val}\) & \(\sigma_{\rm formal}\) &
\(\sigma_{\rm robust}\) & Infl. & Formal cov. & Robust cov. & Formal FPR & Robust FPR\\
\hline
Conservative & Matched & 5000 & 5000 & 9.513e-03 & 9.602e-03 & 1.01 & 0.950 & 0.953 & 0.052 & 0.052\\
Conservative & Stress  & 5000 & 5000 & 9.513e-03 & 1.142e-02 & 1.20 & 0.865 & 0.949 & 0.135 & 0.049\\
Moderate     & Matched & 5000 & 5000 & 4.490e-04 & 4.502e-04 & 1.00 & 0.951 & 0.952 & 0.053 & 0.051\\
Moderate     & Stress  & 5000 & 5000 & 4.490e-04 & 1.360e-03 & 3.03 & 0.438 & 0.950 & 0.550 & 0.048\\
Optimistic   & Matched & 5000 & 5000 & 4.182e-05 & 4.239e-05 & 1.01 & 0.950 & 0.954 & 0.052 & 0.048\\
Optimistic   & Stress  & 5000 & 5000 & 4.182e-05 & 1.972e-04 & 4.71 & 0.317 & 0.949 & 0.685 & 0.051\\
\hline
\end{tabular}
\end{table*}

\begin{table*}[t]
\centering
\caption{\textbf{Julia nonzero-injection bias and RMSE outputs for \((\chi_1,\chi_2)=(10^{-3},-5\times10^{-4})\).}}
\label{tab:app_mc_outputs_b}
\footnotesize
\begin{tabular}{llcccc}
\hline
Scenario & Regime & Bias \(\chi_1\) & Bias \(\chi_2\) & RMSE \(\chi_1\) & RMSE \(\chi_2\)\\
\hline
Conservative & Matched & 1.511e-05 & -1.505e-07 & 1.558e-03 & 9.521e-03\\
Conservative & Stress  & -2.554e-05 & -7.437e-05 & 1.981e-03 & 1.131e-02\\
Moderate     & Matched & -1.245e-06 & -2.746e-06 & 1.723e-04 & 4.509e-04\\
Moderate     & Stress  & -2.921e-06 & 6.279e-06 & 2.761e-04 & 1.366e-03\\
Optimistic   & Matched & -8.895e-08 & 1.038e-06 & 2.748e-05 & 4.170e-05\\
Optimistic   & Stress  & -2.913e-07 & 4.552e-07 & 3.734e-05 & 2.003e-04\\
\hline
\end{tabular}
\end{table*}

\subsection{Process and covariance diagnostics}
\label{app:process_diagnostics}

\begin{figure}[H]
\centering
\includegraphics[width=0.88\linewidth]{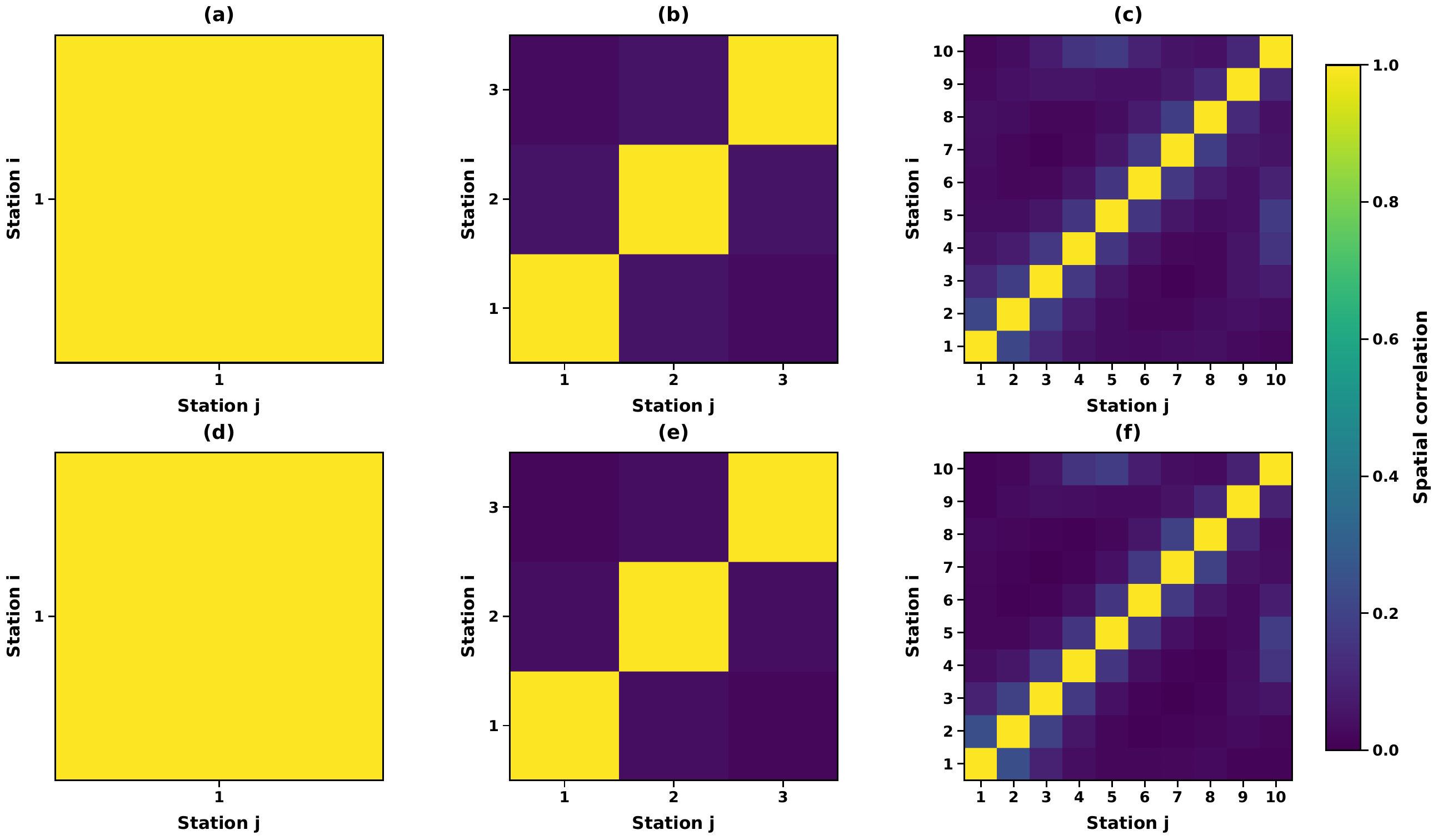}
\caption{\textbf{Fitted and Stress spatial network-correlation matrices.}
(a)--(c) The upper row shows the fitted GLS spatial-correlation matrices for the Conservative, Moderate, and Optimistic networks, respectively. (d)--(f) The lower row shows the corresponding Stress-generation matrices in the same scenario order. Thus (a) and (d) are Conservative, (b) and (e) are Moderate, and (c) and (f) are Optimistic. Off-diagonal entries are determined by station angular separation through Eq.~(\ref{eq:network_spatial_correlation}). These panels show the spatial factor of the separable low-rank covariance, not the complete time-domain covariance matrix.}
\label{fig:app_network_covariance}
\end{figure}

Figure~\ref{fig:app_network_covariance} displays the spatial factor used by the low-rank stochastic network term in Eq.~(\ref{eq:network_spatial_correlation}).

The one-station Conservative case is necessarily the trivial \(1\times1\) self-correlation and therefore contains no empirically separable cross-station structure. In the three-station Moderate network, the fitted off-diagonal correlations are \(0.035\)--\(0.055\), while the corresponding Stress values are \(0.023\)--\(0.043\). The ten-station Optimistic network resolves the expected distance-dependent banding: the strongest displayed off-diagonal pair is \(0.218\) in the fitted matrix and \(0.244\) in the Stress matrix, whereas widely separated pairs fall to approximately \(0.005\)--\(0.04\). The matrices are symmetric, have unit diagonal, and are positive definite in every construction. Their difference is intentional: Stress generation changes the correlation amplitude and length scale, so the injection--recovery test probes covariance misspecification rather than merely repeating the fitted GLS model. Temporal low-rank modes and site-specific AR(1) terms complete the time-domain covariance and are not represented by this heat map.

\begin{figure}[H]
\centering
\includegraphics[width=0.86\linewidth]{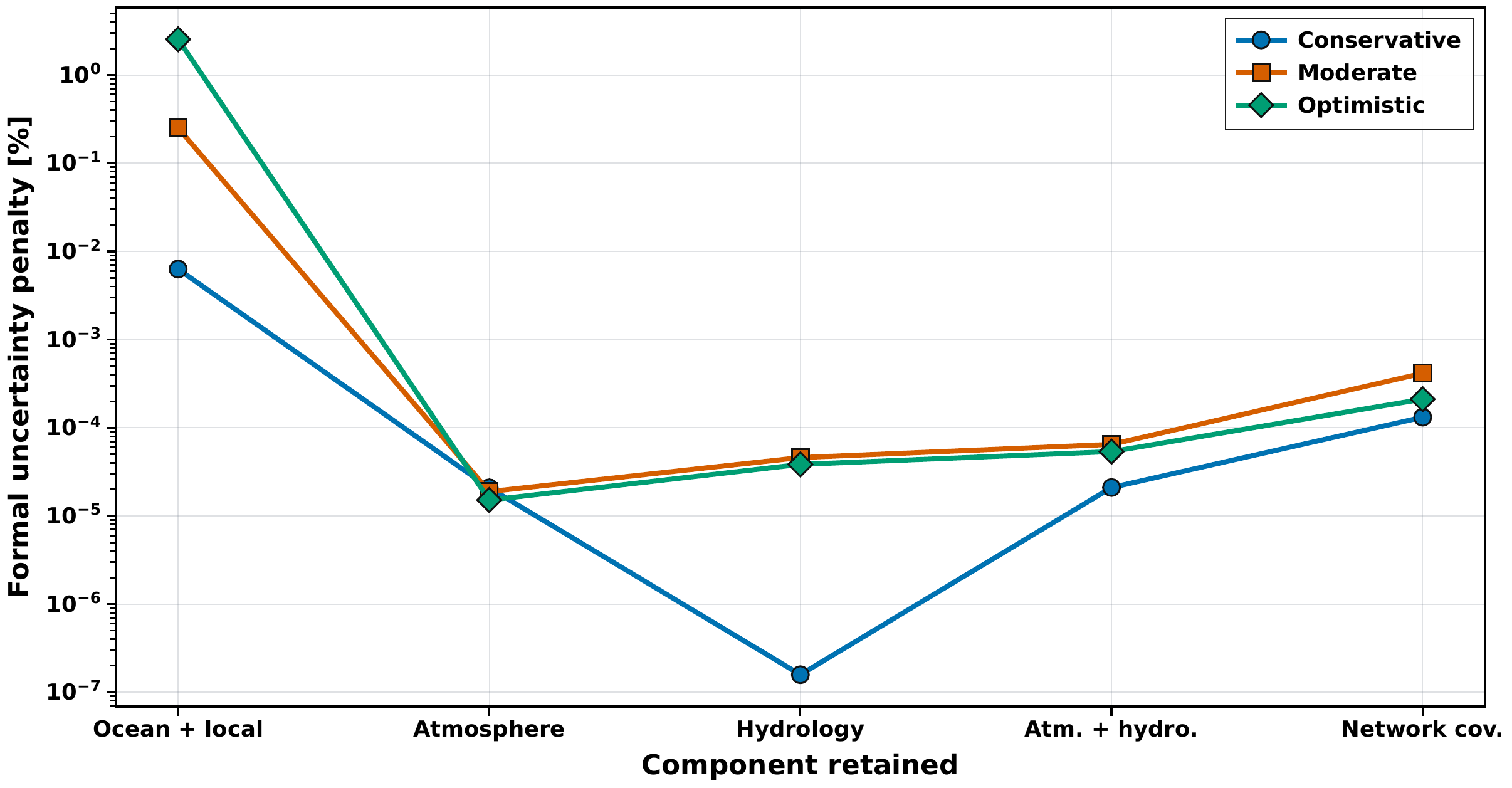}
\caption{\textbf{Formal process-ablation impacts.}
 Each curve compares the full joint-prior forecast with calculations in which the indicated fitted process is removed; the three curves correspond to the Conservative, Moderate, and Optimistic scenarios. All positive values are plotted at their computed magnitudes, without imposing a display floor; the logarithmic axis exposes effects spanning more than seven orders of magnitude. Small formal penalties for atmospheric, hydrological, or network-covariance terms do not establish robustness to their misspecification; that separate issue is assessed by the Stress injection--recovery validation in Figure~\ref{fig:v13_stress_validation}.}
\label{fig:app_ablation}
\end{figure}

Figure~\ref{fig:app_ablation} records process-removal diagnostics. For each process the plotted percentage is \(100(\sigma_{\rm full}/\sigma_{\rm removed}-1)\), so a positive value is the formal uncertainty cost of retaining that process in the full model. The ocean/local-loading block has the largest formal effect, reaching \(2.543\%\) in the Optimistic scenario; the other fitted process blocks have much smaller effects because the formal result remains dominated by the global angular priors.

\begin{figure}[H]
\centering
\includegraphics[width=0.86\linewidth]{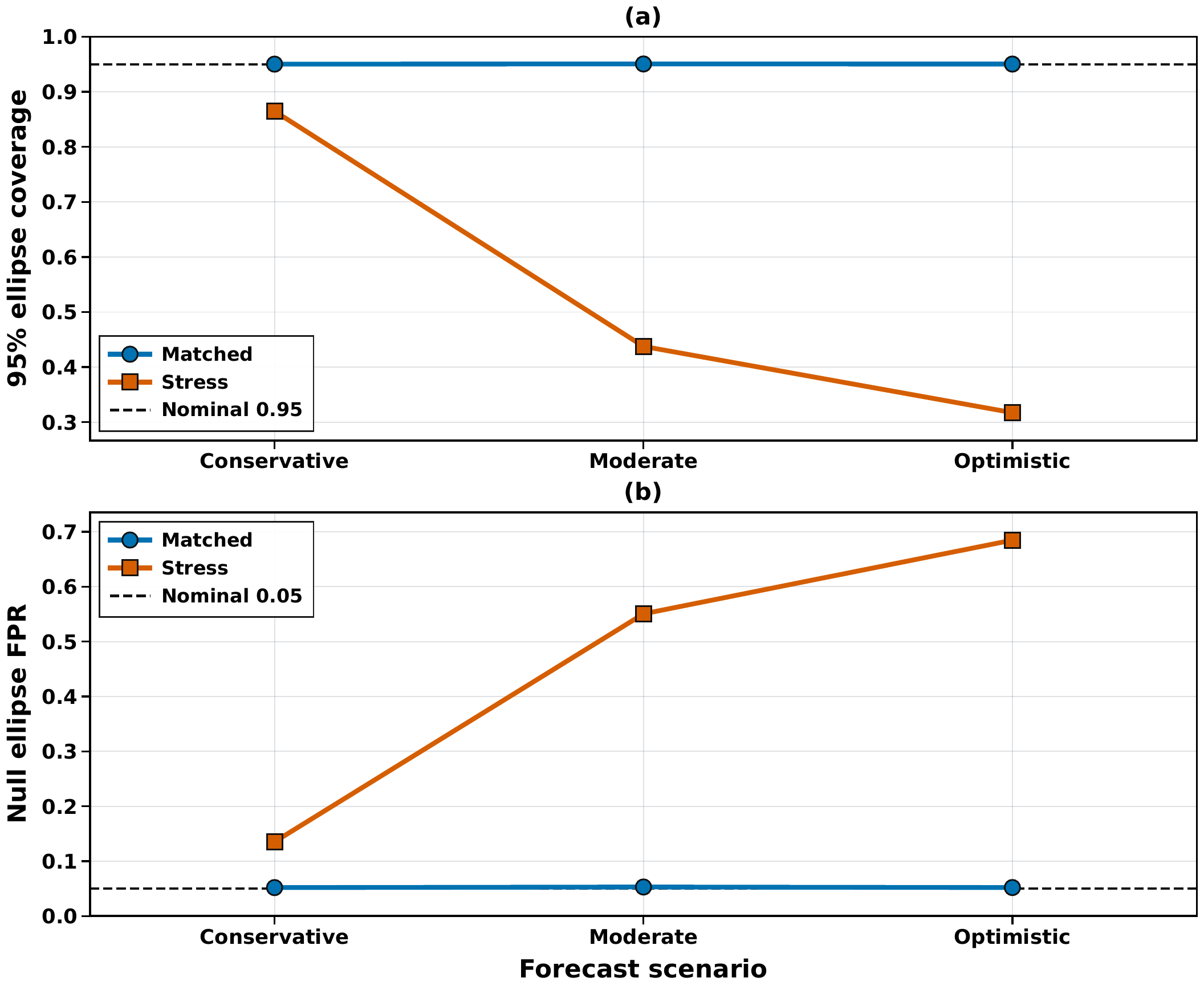}
\caption{\textbf{Formal injection--recovery coverage and false-positive behavior.}
(a) The upper panel shows formal \(95\%\) joint-ellipse coverage for Matched and Stress ensembles. (b) The lower panel shows formal null joint-ellipse false-positive rates. The dashed references are \(0.95\) and \(0.05\), respectively. Robust held-out results are not shown here; they appear in Figure~\ref{fig:v13_stress_validation} and Table~\ref{tab:v13_mc_results}.}
\label{fig:app_injection_recovery}
\end{figure}

The ocean/local-loading penalty rises from \(0.0063\%\) to \(0.251\%\) and \(2.543\%\) across the Conservative, Moderate, and Optimistic configurations. This progression shows that constituent-resolved loading becomes more consequential as the nominal statistical sensitivity improves. By contrast, removing the fitted atmospheric, hydrological, combined environmental, or stochastic network-covariance block changes the formal worst-axis uncertainty by less than \(5\times10^{-4}\%\) in these calculations. Those small formal ablations do not imply that the processes are physically absent or harmless: they quantify local variance changes within the assumed fitted model, whereas omitted amplitudes, phase errors, temporal mismatch, and covariance mismatch are tested separately by the Stress ensembles. The distinction explains why a process can have a small Fisher ablation yet still contribute to the loss of calibration seen in Figure~\ref{fig:app_injection_recovery}.

For completeness, Figure~\ref{fig:app_injection_recovery} shows the uncalibrated formal-covariance behavior in both Matched and Stress regimes. It is retained as a diagnostic complement to the main-text held-out validation figure.

The Matched ensembles remain close to nominal: formal joint-ellipse coverage is \(0.950\), \(0.951\), and \(0.950\), with false-positive rates \(0.052\), \(0.053\), and \(0.052\). Under response-expanded Stress generation, the same formal covariance gives coverage \(0.865\), \(0.438\), and \(0.317\), while the false-positive rates rise to \(0.135\), \(0.550\), and \(0.685\). The increasing separation toward the Optimistic configuration is the expected signature of a forecast becoming limited by the declared mismatch budget rather than by its fitted covariance alone. Figure~\ref{fig:app_injection_recovery} deliberately shows this uncorrected failure mode; Figure~\ref{fig:v13_stress_validation} and Table~\ref{tab:v13_mc_results} then show that covariance estimated on the independent calibration half restores approximately nominal performance on the held-out validation half. The two figures therefore answer different questions and are not duplicate tests.

The numerical records in this appendix are synthetic products of the Julia implementation. They are included to document the exact scenario definitions and the reported output values; they should not be interpreted as measurements of the terrestrial residual environment or as evidence for a nonzero lunar cross-residual.





\end{document}